%% file: main.tex
\documentclass[conference,compsoc]{IEEEtran}
\usepackage{subfigure}
\usepackage{hyperref}
\usepackage{xcolor}
\usepackage{enumitem}
\usepackage{graphicx}
\usepackage{amsmath}
\usepackage{amsfonts}
\usepackage{booktabs}
\usepackage{diagbox}
\usepackage{multicol}
\usepackage{multirow}
\usepackage{cleveref}
\usepackage{ulem}
\usepackage{bm}

\crefname{figure}{\textbf{Fig.}}{\textbf{Figs.}}
\crefname{table}{\textbf{Table.}}{\textbf{Tables.}}
\crefname{equation}{\textbf{Eq.}}{\textbf{Eqs.}}
\crefname{section}{\textbf{Sect.}}{\textbf{Sects.}}

\ifCLASSOPTIONcompsoc
  \usepackage[nocompress]{cite}
\else
  \usepackage{cite}
\fi
\ifCLASSINFOpdf
\else
\fi
\begin{document}
\title{
% Bare Demo of IEEEtran.cls for\\ IEEE Computer Society Conferences
% DMTG: An Open-world Mouse Trajectory Generator Based on\\ Diffusion Networks Against the Commercial CAPTCHAs
% 基于熵控DDPM的类人鼠标轨迹仿真机器人
% DMTG: Humanoid Mouse Trajectory Generation Bot\\ Based on Entropy-Controlled Diffusion Networks
DMTG: A Human-Like Mouse Trajectory Generation Bot \\ Based on Entropy-Controlled Diffusion Networks
}


%%% 作者
\author{
\IEEEauthorblockN{Jiahua Liu}
\IEEEauthorblockA{University of Michigan, USA\\
Email: jiahual@umich.edu}
\and
\IEEEauthorblockN{Zeyuan Cui}
\IEEEauthorblockA{Email: zeyuancui@hotmail.com}
\and
\IEEEauthorblockN{Wenhan Ge}
\IEEEauthorblockA{College of Computer Science\\
Sichuan University, China\\
Email: saltriangle@163.com}
\and
\IEEEauthorblockN{Pengxiang Zhan}
\IEEEauthorblockA{College of Computer Science\\
Sichuan University, China\\
Email: hanpengxiang@stu.scu.edu.cn}
}


% compsoc conferences compared to traditional conferences), use this
%Telephone: (800) 555--1212, Fax: (888) 555--1212}


\maketitle

\begin{abstract}
    CAPTCHA protects normal business and prevents resource misuse and data theft by distinguishing between human operations and machine programs in the network. Advances in machine learning techniques have made traditional image-based and text-based CAPTCHAs susceptible to targeted attacks, thus modern CAPTCHAs such as GeeTest and Akamai incorporate behaviors such as mouse track detection for senseless detection.
    Present methods for bypassing mouse behavior detection mainly implement simulation generation of mouse trajectories through representation learning, adversarial learning, or reinforcement learning. Such methods suffer from the difficulty of fully mimicking human style and the lack of all-round testing, which makes it challenging to evaluate the effectiveness of anti-bot measures. 
    To address these issues, this paper proposes a mouse trajectory generation and evaluation framework based on an entropy-controlled diffusion model: DMTG, which accepts complexity control and generates mouse control curves approximating real human samples. The framework also provides both white-box and black-box testing methods for verifying whether the model can have the ability to obfuscate human-machine operations and pass the online commercial CAPTCHA test.
    The experiments are centered around black-box and white-box tests. In the white-box tests, DMTG has a significantly lower probability of being detected as a bot than the other models, as evidenced by an average 4.75\% to 9.73\% reduction in the bot recognition accuracy, and 5.11\% to 10.51\% in macro F1s. Additionally, we evaluated the performance of the DMTG framework on commercial CAPTCHA systems to assess its effectiveness in real-world scenarios. Due to copyright and proprietary restrictions, detailed results are not provided in this paper. Furthermore, DMTG can effectively mimic the physical properties of human operations, including slow initiation and differences in applied force for different directions. The results show that DMTG is more likely to generate mouse trajectories in human expectations and can provide quantitative simulation metrics.
\end{abstract}

\input{secs/intro}
\input{secs/relwk}
\input{secs/mtd}
\input{secs/exp}
\input{secs/concul}



\bibliographystyle{IEEEtran}
\bibliography{refs}


% \end{thebibliography}

\input{secs/apedx}

\end{document}

%% file: secs/intro.tex
\section{Introduction}\label{sec:intro}

CAPTCHA (Completely Automated Public Turing test to tell Computers and Humans Apart) \cite{CAPTCHA} distinguishes human and automated mechanical operations through complex logic to prevent large-scale data theft and misuse. Traditional CAPTCHA methods are based on image \cite{img-captcha-1,img-captcha-2}, text \cite{txt-captcha-1,txt-captcha-2,txt-captcha-3}, and other medias. Due to the development of Artificial Intelligence (AI), these CAPTCHAs can be easily bypassed through large-scale supervised learning. For example, CNN \cite{img-captcha-1} or YOLO \cite{YOLO} recognition can locate image elements. Large Language Models (LLM) \cite{llm-captcha-1} are good at reading comprehension or solving logical judgments. As a result, many advanced CAPTCHA advocate developing invisible detection, such as the mouse or finger trajectory or other behaviors.

1) Style Variation: Existing methods face challenges in achieving controllable random generation of mouse trajectories that cater to the distinct styles of different operators. 2) Real-World Applicability: Current approaches primarily rely on white-box testing and quality evaluation metrics like accuracy, lacking sufficient experimentation in large-scale real-world CAPTCHA black-box environments. This limitation weakens the robustness and generalizability of most methods.

Since dual unknowns of the data and the model, the mouse trajectory behavior detection methods for most commercial CAPTCHAs are in a black box. 

Existing methods for obfuscated generation of mouse trajectories mainly include adversarial or reinforcement learning \cite{mt-becaptcha, mt-captcha-rl-1, mt-captcha-rl-2, cc-net}, and encoder-decoder \cite{mt-SapiAgent, T-Detector}. Most of these methods rely on simulated adversarial environments in laboratory settings and lack evaluations in real-world network scenarios. Their issues primarily include the following two points. 1) \textbf{Anthropomorphic Stylization}: Existing methods struggle to generate mouse trajectories that are tailored to different web pages and user styles, making them easy to track and trace; 2) \textbf{Practical Testing Limitation}: Current approaches primarily rely on white-box testing and evaluation metrics like accuracy, lacking sufficient black-box experimentation of large-scale real-world CAPTCHA environments. This limitation weakens the robustness and generalizability of most methods.

We propose DMTG (Diffusion-based Mouse Trajectory Generator), an end-to-end approach for generating realistic mouse trajectories based on the theory of entropy-controlled DDIM \cite{DDIM}. Entropy-controlled DDIM adjusts the local uncertainty of generated mouse trajectories by an energy coefficient, allowing the trajectories to exhibit human-like randomness while achieving controlled objectives. Compared to prior works, the primary innovation of DMTG lies in its ``\textbf{controllable randomness}". On the one hand, due to the diffusion process, identical web pages yield differing trajectories in each generation turn, which adapts to humans' unpredictability. On the other hand, the constraints imposed by the style transfer module enable DMTG to mimic and clone the user's historical behaviors relatively, thereby blurring the distinctions between human and machine operations.

The main contributions of this paper are as follows: 
\begin{itemize}
    \item We introduce the DMTG framework, which can generate highly realistic mouse trajectories to bypass the CAPTCHA detectors.
    \item We improve the DDIM model using information entropy theory, enabling users to control trajectory length, complexity, and target position.
    \item We propose white-box testing methods including distribution measurement and deep learning prediction to assess the fundamental differences between simulated mouse trajectories and human activities.
\end{itemize}

The rest of the paper is structured as follows: \cref{sec:relwk} presents the current approaches of CAPTCHA and Generative Artificial Intelligence (GAI). \cref{sec:mtd} describes the modeling and methodology of DMTG. \cref{sec:exp} discusses the experiments and interpretations. \cref{sec:concul} summarizes the whole paper and mentions our future works. 

%% file: secs/relwk.tex
\section{Related Work}\label{sec:relwk}

% 本章将介绍现有的CAPTCHA技术方案和生成式人工智能的研究现状，用于提供DMTG的知识基础和对比方法。
This section will present the current state of research on existing CAPTCHA technology schemes and GAI, which provide the knowledge base and comparison methodology for DMTG.

\subsection{CAPTCHA and Deceivers}\label{sec:relwk-captcha}

Existing CAPTCHA \cite{CAPTCHA} can be divided into four basic categories: text-based, image-based, audio-based and user behavior-based. Text-based CAPTCHA verifies whether the user has entered the correct text string using distorted or noisy text images to distinguish between automated programs and real humans \cite{txt-captcha-1,txt-captcha-2,txt-captcha-3}. Image-based CAPTCHA synthesizes the logical differences between humans and machines through image semantic understanding \cite{img-captcha-1,img-captcha-2}. For example, Tencent's VTT\cite{tencent-vtt} uses style migration in conjunction with 3D shapes to generate check images. Google's reCAPTCHA\cite{google-recaptcha} uses adversarial perturbations to interfere with AI's recognition of images to distinguish humans from bots. CAPTCHA based on phonetic or biological information uses voiceprint or timbre to differentiate between human and synthetic speech \cite{audio-captcha-1}, which is less common than images and text.

CATPCHAs based on images, text, and sounds are significant because these verification methods can provide an obvious target for the screening object to process. The AI robot can be purposefully simulated to bypass these filters because of their strong willingness \cite{easy-not-secure}. For example, Deng et al. \cite{llm-captcha-1} exploited the cross-domain logic capabilities of LLMs to bypass image understanding tasks. Hossen \cite{recaptcha-yolov3} uses YOLOv3 \cite{yolov3} to locate elements in reCAPTCHA v2. Wang \cite{txt-rcnn-att-chinese} prefers the word obfuscation-generating units combined with Recurrent Neuronal Networks (RNN) \cite{RNN}, Convolutional Neural Networks (CNN) \cite{CNN}, and attention mechanisms \cite{ATT} to achieve Chinese validator bypass. Due to the vulnerability of explicit CAPTCHA to attacks, CAPTCHA based on implicit user behavior verification has been chosen as a new type of verifier.

The implicit nature of the user behavior CAPTCHA is reflected in the black-box nature, including the unnoticeable, unknowable, and unrepeatable attributions in detection. Mouse2Vec \cite{Mouse2Vec} uses the Transformer self-supervised framework for learning semantic representations of mouse behavior. ReMouse \cite{ReMouse} provides novel real-world mouse dynamics datasets and finds that the irreducible nature of human manipulation can be used to circumvent replayed machine simulation behavior. Acien\cite{phone-traj} uses RNN for gesture recognition to realize user touch behavior detection of smartphones. Jin\cite{jin-mouse-traj} uses CNN and RNN to distinguish human and machine mouse trajectories. T-Detector\cite{T-Detector} uses RNN and CNN to model human-computer operational differences in gaming scenarios and finds that humans have more ineffective activities. As a long-term detection process, the mouse track CAPTCHA is considered to have a certain degree of review and correction capabilities. This makes the attack robots facing it relatively more complicated and requires considering the accumulation of multiple factors.

The invisibility and complexity of user behavior CAPTCHAs make mouse trajectory attacks in their infancy. Existing methods focus on using distribution fitting to generate behaviors of fake humans to bypass the detection. BeCAPTCHA-Mouse\cite{mt-becaptcha} uses Generative Adversarial Networks (GAN) \cite{GAN2} to generate mouse trajectory distributions that mimic those of humans. Folch\cite{mt-captcha-3} uses a hybrid mechanism of multiple straight lines and curves to simulate and detect the robot mouse operation trajectory. SapiAgent \cite{mt-SapiAgent} uses an auto-encoder to simulate human operations. CC-Net \cite{cc-net}designed a set of image- and text-based multimodal network-generated basic sequences of computer operations and outperformed on the MiniWob++ \cite{MiniWob} test. Tsingenopoulos \cite{mt-captcha-rl-2} designed an automated web browser based on Reinforcement Learning (RL) \cite{RL} to bypass the behavioral scoring mechanism of reCAPTCHA v3, and found that if humans are involved in initiating the browsing, the subsequent non-human operating agent is extremely difficult to detect.

Summarily, among all CAPTCHAs, user behavior detection can provide the website with the greatest protection against machine abuse. Although the existing solutions to bypass user behavior detection mostly use adversarial generation to promote robots to imitate humans, there are still some problems in track generation: 1) \textbf{Anthropomorphic Stylization}. The non-repeatability of human operations found by ReMouse \cite{ReMouse}, the ineffective human operation characteristics found by T-Detector \cite{T-Detector}, and the human hot-start bypassing found by Tsingenopoulos \cite{mt-captcha-rl-2}, all show that there is still room for improvement in existing mouse track generation schemes. Thus, the new type of mouse trajectory generation needs to imitate the randomness of human operations and the historical styles of different humans to form an attack scheme with inconspicuous differences. 2) \textbf{Practical Testing Limitationn}. Although some solutions have passed the reCAPTCHA experiment, they still need to be more widely verified in a variety of commercial CAPTCHAs.

\subsection{Generative Artificial Intelligence}

GAI creates creative, hard-to-code, clear, persuasive, and relatively versatile text or visual output based on large amounts of training data, augmenting the capabilities of existing human workers by increasing productivity and bringing greater efficiency to repetitive editing tasks \cite{GAI-product-effect}. Modern GAI tasks mainly focus on fields like sequences (video, text, audio), images/graphics, and chemistry/biomedicine, aiming to generate working modes or result distributions that are close enough to humans \cite{GAI-defination}. These tasks mainly rely on four frameworks: Encoder-Decoder \cite{AutoEncoder, VAE, DAE, OpenVoice}, GAN \cite{GAN2, GAN, CycleGAN, StyleGAN}, Transformer \cite{ATT, ViT, GPT, Dall-E}, and Diffusion Network \cite{DiffusionNet, DDPM, AlphaFold3}. These four architectures can be grouped according to the global distribution and local context of modeling. Among them, Encoder-Decoder and GAN are typical global sampling, which advocate representing the input as an independent distribution and resampling it as new data. While Transformer and Diffusion Networks follow local context sampling, which cumulatively obtains new data whose elements are related to each other through time-step changes.

As discussed in \cref{sec:relwk-captcha}, existing mouse trajectory generation methods mainly employ GANs and Encoder-Decoder architectures. These architectures are prone to biases in local features, resulting in identification as robotic behavior. And we believe that Transformer and Diffusion Networks can correct such local disparities. Therefore, we will focus on discussing the current state of Transformer and Diffusion Networks here.

Although both Transformer and Diffusion Networks reason through context, there is a dimensional difference. Transformer performs contextual vacancy reduction more through spatially consistent logic. For example, ViT \cite{ViT} and DALL-E \cite{Dall-E} can generate scene-dependent images from the positional encoding and pixel content of image patches. GPT \cite{GPT} and LLaMA \cite{llama} are good at generating subsequent words with the help of conditional transfer probabilities from the preceding content of the text. While Transformers demonstrate efficient information-gathering capabilities, they still pose risks of forgetting in ultra-long, infinite, or cross-domain scenarios \cite{vit-gan-survey}. At the same time, because Transformer's original models are generally large and complex, it is difficult to balance its computational efficiency and accuracy \cite{vit-survey}. These requirements constrain the deployment of Transformer to low-resource computing devices, making the computational cost difficult to control.

Diffusion Network generates special results from noise by utilizing thermodynamic diffusion principles and their inverse processes on time slices \cite{DiffusionNet}. Compared to the Transformers, Diffusion Networks better represent the continuity of frames and states. For example, DDPM \cite{DDPM} implements a deep diffusion model fitting method for generating high-quality images of the same subject. AlphaFold3 \cite{AlphaFold3} utilizes a diffusion model to generate the folded spatial structure of a protein macromolecule. MedSegDiff \cite{MedSegDiff} has designed a DDPM-based U-Net \cite{u-net} architecture for segmenting regions in medical images. UniDiffuser \cite{UniDiffuser} presents a multi-distributed diffusion model for generating realistic text or images that rival customized large-scale Transformer capabilities while maintaining efficiency, demonstrating the potential of Diffusion Networks. Current challenges in Diffusion Networks focus on the distribution, direction, process, and feedback controllability of generation \cite{difu-survey, v-difu-survey, difu-survey-2} so that researchers still have room for improvement in locally customized generation.

%% file: secs/mtd.tex
\section{Methodlogy}\label{sec:mtd}

\subsection{Issue Definitions}

As discussed in \cref{sec:intro,sec:relwk-captcha}, DMTG primarily revolves around two objectives: \textbf{Anthropomorphic Stylization} and \textbf{Practical Testing Limitations}, in realistic mouse trajectories generation. Anthropomorphic stylization necessitates GAI to mimic human habits, demeanor, and even errors in trajectory generation. Practical testing limitation requires the generated mouse trajectories to circumvent generic commercial-grade CAPTCHA detectors. To address these, we need to tackle three fundamental Research Questions (RQ): 
\begin{enumerate}[font={\bfseries},itemindent=1em,label={RQ\arabic*}]
    \item How to construct mouse trajectories that are simultaneously compatible with randomness and purposiveness?\label{rq1}
    \item How to ascertain that such trajectories adhere to human operational styles, making them indistinguishable from those of the other bots?\label{rq2}
    \item How to evaluate the model's ability to bypass commercial CAPTCHAs?\label{rq3}
\end{enumerate}

Our method consists of the following main 4 steps:
\begin{enumerate}[font={\bfseries},itemindent=1em,label={Step-\arabic*.}]
    \item \textbf{Data collection}: we use the SapiMouse dataset \cite{SapiMouse} and the Open Images V7 dataset \cite{openimages} as our training datasets, which contain samples of human mouse shenanigans.
    \item  \textbf{Model Construction}: the goal of DMTG is to mimic human trajectories, and thus RQ1 and RQ2 need to be solved. The RQ1 is solved by the main body of the DMTG's entropy-controlled diffusion networks named $\alpha$-DDIM. And RQ2 is controlled by the DMTG's unique loss function. Both RQ1 and RQ2 requires that the $\alpha$-DDIM be able to generate directional trajectories from random seeds, implying that the trajectory $\hat{Y}$-generation function: \begin{equation}\label{eq:f}
        \hat{Y}=f(\vec{X},\epsilon, \alpha, m)
    \end{equation} needs to have 4 inputs, the target directions $\vec{X}$ of the start and end of the mouse movement, the random initial noise $\epsilon$, the  complexity control factor $\alpha$ to terminate the diffusion progress, and the node num $m$ that control the action frames of the whole trajectory.
    \item \textbf{System Development}: $\alpha$-DDIM only realizes the core functionality of the DMTG system, in order to realize a usable system, it is necessary to add: fingerprint assembler, peripheral controllers, and other supporting facilities.
    \item \textbf{Evaluation}: we designed black and white box tests separately. The white box identifies authentic trajectories by constructing a fully supervised AI model to evaluate RQ2. While the black box testing utilizes commercial CAPTCHA to test bypass rates to answer RQ3.
\end{enumerate}

The remaining sub-sections of this section will illustrate how DMTG is implemented and address the posed RQs.

\subsection{Data Collection}

The dataset used by DMTG is a mixture of the SapiMouse dataset \cite{SapiMouse} and the Open Images V7 dataset \cite{openimages}. SapiMouse collected human mouse trajectories from 120 participants, where the data is in the form of a set of coordinates along with corresponding timestamps containing mouse events. Open Images V7 provides annotations for semantic understanding of images. We utilize the sequences of mouse coordinates under the ``traces" label in Open Images. After combining these coordinate data, we randomly sampled 1 million instances as human samples to train the robot's modeling capability.

\subsection{Model Construction}\label{sec:mtd-dmtg}

\begin{figure*}
    \centering
    \includegraphics[width=\linewidth]{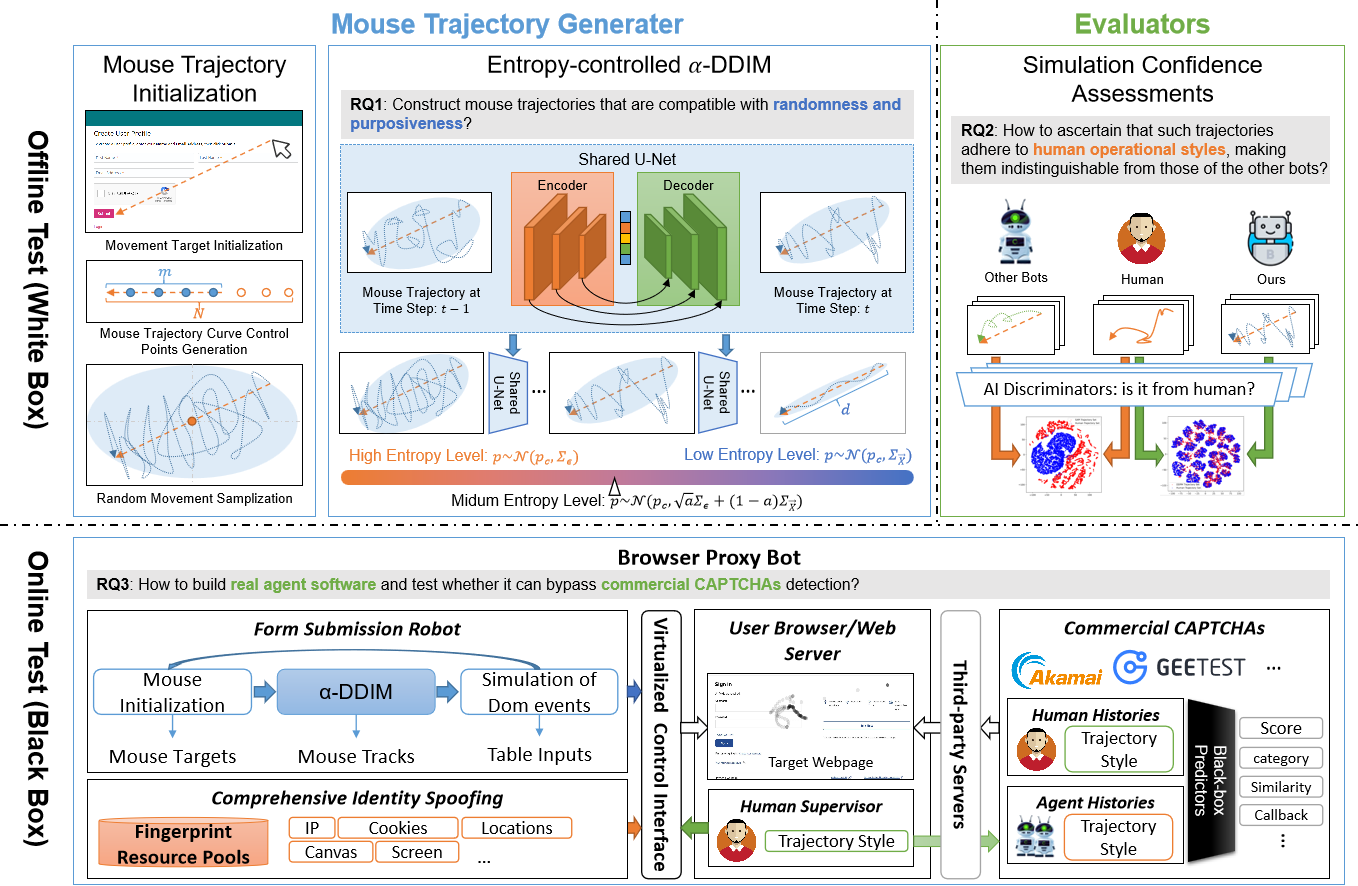}
    \caption{DMTG Framework}
    \label{fig:framework}
\end{figure*}

The DMTG framework has two components: a generator for generating controlled random trajectories, and a series of raters for evaluating bypassing and imitation capabilities. The structure of the whole framework is shown in \cref{fig:framework}. The generator consists of two main functions, the initialization to formulate the mouse agent task, and the $\alpha$-DDIM to generate ``controlled random” trajectories in RQ1. Besides, we design a Browser Proxy Bot for simulating real operating environments and deployed the $\alpha$-DDIM model. The quality evaluators also have two components. The Simulation Confidence Assessments answer the human imitation ability assessment in RQ2, and The Commercial CAPTCHA Validations measure the generalizability and real-world applicability in RQ3.

Suppose an arbitrary mouse coordinate is $p=\langle x,y\rangle~s.t.~x\in [0,W), y\in [0,H)$, where $W$ is the width of the screen and $H$ is the height of the screen. Then a mouse track with $m$ moves can be represented as a set of mouse coordinates with $m+1$ nodes: $\{p\}_{m+1}=\{\langle x_0,y_0\rangle,\langle x_1,y_1\rangle,\cdots, \langle x_m,y_m\rangle\}$. In this set, $p_0=\langle x_0,y_0\rangle$ is the initial position of the mouse and $p_m=\langle x_m,y_m\rangle$ is the final position. Thus the $\vec{X}$ can be expressed as $\vec{X}=\{p_0,p_m\}$.

For a mouse control that is specified to move $m$ times, a simple approach is to move directly from $\langle x_0,y_0\rangle$ through the same $m$ displacements in the direction of $\vec{X}$, with each displacement $\frac{\langle x_0,y_0\rangle-\langle x_m,y_m\rangle}{m}$. A complex method is to apply completely random coordinates for the first $m-1$ turns of the $m$ displacements and then move directly to $\langle x_m,y_m\rangle$ for the last one. However, whether it is the simple or the complex one, it reveals that an automated program controls the mouse because the track does not conform to human operating habits. So we need to make the mouse trajectories between the two, to achieve a relative balance between purposefulness and randomness. 

Based on this goal, it was easy to think of using DDIM \cite{DDIM} for generating mouse trajectories. DDIM is based on the inverse process of the thermodynamic entropy increase phenomenon and can generate all the transition states and ways of changing from completely random to direct utilizing neural networks. Thus, DDIM can generate human-like states at some point, allowing DMTG to generate highly realistic mouse trajectories. However, existing DDIMs fail to meet our requirements for controlling curve complexity. Therefore, we need to modify the model accordingly to archive $\alpha$-DDIM.

\subsubsection{Mouse Trajectory Initialization}

In the initialization phase we need to complete the initialization of the three base parameters in \cref{eq:f}. Since each generation process of the $\alpha$-DDIM is a stepwise denoising process, the initial input to the model should be a randomized mouse trajectory. We use normal random as our noise initialization function:$\epsilon\sim \mathcal{N}(0,\Sigma_\epsilon)$, where $\Sigma_\epsilon$ is its standard deviation which is given by \cref{eq:epsilon_c}. 

The point $p_c=\langle x_c,y_c\rangle=\langle\frac{x_0+x_m}{2},\frac{y_0+y_m}{2}\rangle$ here denotes the midpoint between the starting point $\langle x_0,y_0\rangle$ and ending position $\langle x_m,y_m\rangle$. Here $k_c$ is the scaling factor. According to the density function of the normal distribution, 3 times the covariance can make the probability of being within the screen 99\%. Therefore, the scaling coefficient $k_c$ is generally taken to be $\frac{1}{6}$.

\begin{equation}
    \begin{aligned}
        \Sigma_\epsilon = ({(k_c(x_m-x_0))}^2 + {(k_c(y_m-y_0))}^2)\left[\begin{array}{cc}
            1 & 0 \\
            0 & 1
        \end{array}
        \right]
    \end{aligned}
    \label{eq:epsilon_c}
\end{equation}

Since DDIM network cannot handle continuous inputs, necessitating downsampling into discrete points. Also due to the limitation of the number of sampling points $m$ in \label{eq:f}, the final input $X_R$ is defined by \cref{eq:masked_input}. 

\begin{equation}
    \begin{aligned}
        X_R =& \{\langle x_0,y_0\rangle\} \Vert \{p_c+\epsilon_i\}_{i=1}^{m}
            \Vert \{\langle x_m,y_m\rangle\} \Vert \{\langle 0,0\rangle\}_{N-m}
    \end{aligned}
    \label{eq:masked_input}
\end{equation}
where $0<m\leq N$ represents the generated length, and $\Vert$ denotes the concatenation symbol. $N$ is the max length limited by models. Simultaneously, to ensure that the input length remains unchanged, we will append $N-m$ masks $\langle 0,0\rangle$ in the subsequent steps.

\subsubsection{$\alpha$-DDIM Framework}

First, we need to understand that there are 2 extreme phases within the original DDIM generation cycle. One is complete randomness and the other is complete order. Complete randomness occurs at the initialization stage of the mouse trajectory, when the points in the curve $X_R$ are all independently Gaussian-distributed sampling points, such as \cref{eq:epsilon_c}. Complete order occurs at the decoding endpoint of the DDIM, when all points in the curve $X_R$ tend to line up in the vector direction of $\vec{X}$, forming a set of straight line segments, e.g., \cref{eq:epsilon_a}. Neither of these extremes is desirable; what we want is a “controlled randomness”, somewhere in the middle of the DDIM generation cycle, that is neither too orderly nor too chaotic. Therefore, we can use a mixture of both Gaussian distributions to represent the sampling, e.g., \cref{eq:p_a}.

\begin{equation}
    \begin{aligned}
        & p_\alpha = p_c + \epsilon_\alpha \\
        s.t.~& \epsilon_\alpha \sim \mathcal{N}(0, a\Sigma_\epsilon+(1-a)\Sigma_{\vec{X}}), 1\geq a\geq 0
    \end{aligned}
    \label{eq:p_a}
\end{equation}

\begin{equation}
    \begin{aligned}
        & \Sigma_{\vec{X}} = \left[\begin{array}{cc}
            {(k_c(x_m-x_0))}^2 & k_c^2(x_m-x_0)(y_m-y_0) \\
            k_c^2(x_m-x_0)(y_m-y_0) & {(k_c(x_m-x_0))}^2
        \end{array}
        \right]
    \end{aligned}
    \label{eq:epsilon_a}
\end{equation}

\begin{equation}
    \begin{aligned}
        q_\sigma ({p_a}|\vec{X}) &:= \mathcal{N}(p_c; a\Sigma_\epsilon, (1-a)\Sigma_{\vec{X}}) \\
         &\simeq \mathcal{N}(p_c; \sqrt{a}\Sigma_\epsilon, (1-a)\Sigma_{\vec{X}}) \\
         &= \mathcal{N}({p_a}; \sqrt{a}\Sigma(\vec{X}), (1-a)\Sigma(X_R))
    \end{aligned}\label{eq:ddim}
\end{equation}

It is worth noting that the human-like trajectory we need in \cref{eq:epsilon_a}, can be approximated by the original process of DDIM \cite{DDIM}, with the inference process shown in \cref{eq:ddim}. Where, the required $a$ serves the same purpose as DDIM's $\alpha_t$. $\Sigma(\cdot)$ is the function that calculates variance. The status of ${X}$ is the same as the original requirement of $x_0$ in DDIM, both symbolizing the structured data; while the role of $X_R$ is analogous to that of DDIM's $I$, representing complete noises.

Based on the assumption that the information entropy $H(\{p_a\})$ of the set of points ${p_a}$ of any controlled curve can be calculated as \cref{eq:h_t}.

\begin{equation}
    \begin{aligned}
        H(\{p_a\}) &= \frac{1}{2}\log((2\pi e)^2|a\Sigma_\epsilon+(1-a)\Sigma_{\vec{X}}|) \\
        & = \log(2\pi e) + \frac{1}{2}\log(|a\Sigma_\epsilon+(1-a)\Sigma_{\vec{X}}|) 
    \end{aligned}\label{eq:h_t}
\end{equation}

Although, we tend to control the entropy of the curve output from the model to control the complexity of the trajectory, this may not be easy. This is because, although there is a constraint on the mixing parameter $a$ of the mixed Gaussian distribution, the value of this $a$ cannot be confirmed, resulting in $H(\{p_a\})$ not being available. We therefore thought of replacing the computation of $H({p_a})$ by utilizing an approximate solution that can be used to reflect the complexity of the curve in the iteration. 

We do this by utilizing the total length of the curve as a replacement. In fact, there is theoretical support for this approach. For the total length of a curve, it is always greater than or equal to the length of the Minimum Spanning Tree (MST) of the point set ${p_a}$. At the same time, controlled randomness also requires that the mouse trajectory be relatively smooth, so that the total length of the curve is ideally equal to the length of the MST as in \cref{eq:l_pa}.

\begin{equation}\label{eq:l_pa}
    L(\{p_a\}) = \sum_{i=1}^{m}\Vert p_a^{(i)}-p_a^{(i+1)}\Vert_2 \simeq L_{MST}(\{p_a\})
\end{equation}

\begin{figure}
    \centering
    \includegraphics[width=0.9\linewidth]{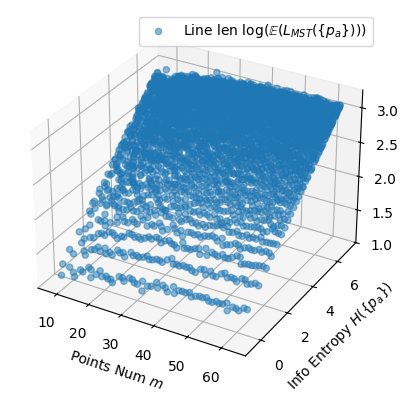}
    \caption{Simulation Results of Points Num $m$, Info Entropy $H(\{p_a\})$, and MST Length $\log(\mathbb{E}(L_{MST}(\{p_a\})))$}
    \label{fig:ddpm-h-a-l-2}
\end{figure}

The logarithm of the mathematical expectation of MST length is positively related to $H(\{p_a\})$. The mathematical derivations are shown in \cref{eq:l_mst} where $\beta$ is a constant associated with $|a\Sigma_\epsilon+(1-a)\Sigma_{\vec{X}}|$.
Our simulation results in \cref{fig:ddpm-h-a-l-2} prove this point.

\begin{equation}
\begin{aligned}\label{eq:l_mst}
    & \log(\mathbb{E}(L_{MST}(\{p_a\}))) \\ 
    & \simeq \log(\beta \sqrt{m|a\Sigma_\epsilon+(1-a)\Sigma_{\vec{X}}|}) \\
    & = \log(\beta) + \frac{1}{2}\log(m) + \frac{1}{2}\log(|a\Sigma_\epsilon+(1-a)\Sigma_{\vec{X}}|) \\
    & = H(\{p_a\}) + \log(\beta) + \frac{1}{2}\log(m) - \log(2\pi e)
\end{aligned}
\end{equation}

Based on the above arguments, we can basically conclude that the controlled randomness we want is related to the total length of the mouse trajectory. Therefore we will input the length of the curve we wish to form normalized as $\alpha$ coefficients into $\alpha$-DDIM to participate in the complexity control of the curve. The diffusion process of the DDIM stops and the output of the generated trajectory is output when, and only when, the curve length normalized at a certain point in time is equal to the preset $\alpha$. These processes can be obtained by adding $\alpha$ constraints to the original formulation of the DDIM.

\begin{figure}
    \centering
    \includegraphics[width=\linewidth]{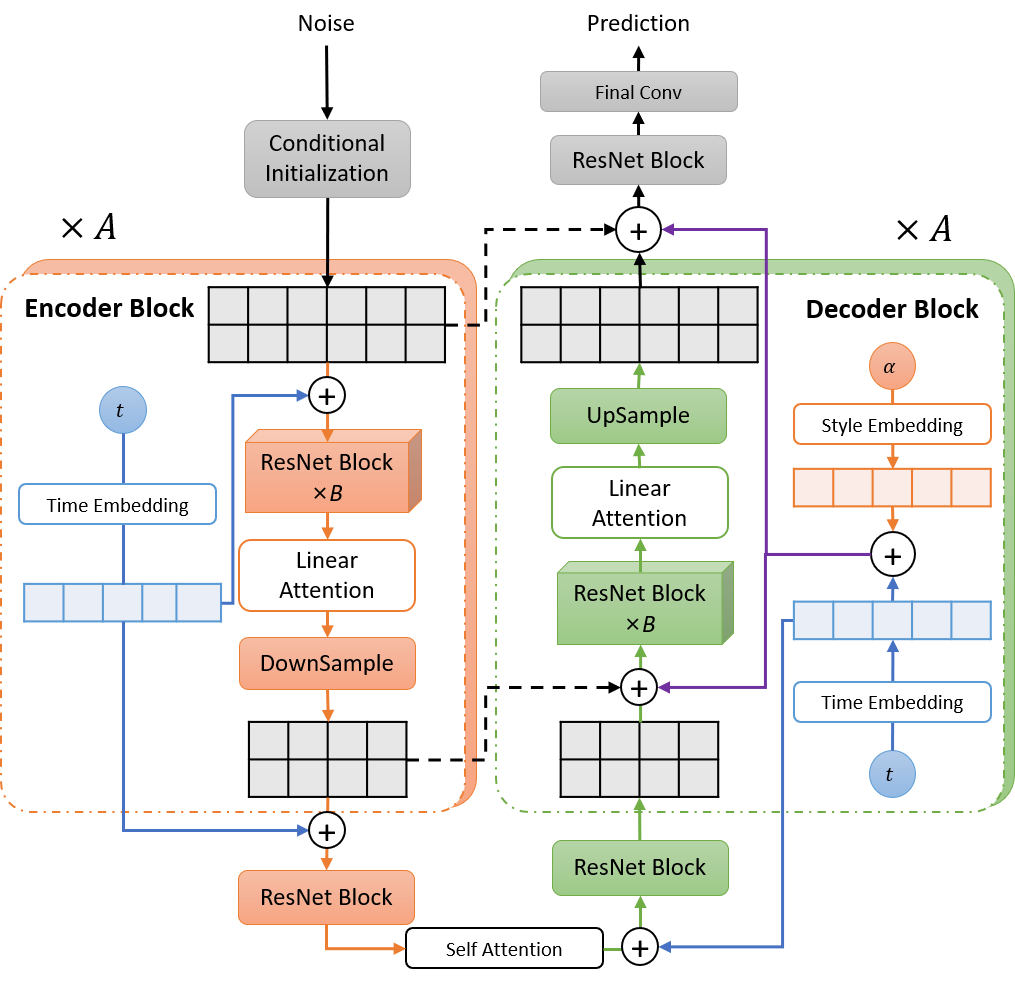}
    \caption{U-Net Model for $\alpha$-DDIM}
    \label{fig:unet}
\end{figure}

We employ the U-Net framework to implement $f_\theta(X_{t-1}|X_t,\alpha)$. The architecture of this model is depicted in \cref{fig:unet}, comprising $2\times A$ layers of encoding and decoding units. In this architecture, the decoding units are influenced by the time step $t$ and the complexity control coefficient $\alpha$ to encode high-dimensional vectors. In contrast, the encoder layers are solely controlled by the time step $t$ to control the iteration rounds. Furthermore, residual connections exist between encoding and decoding layers of the same dimensions to prevent gradient vanishing.

The time embedding in the model utilizes positional encoding similar to the Transformer model \cite{ATT}. The style encoding (\cref{eq:style-emb}) corresponds to the complexity control coefficient $\alpha$. Where, $1/(\alpha+1)$ is responsible for transforming the range of $\alpha$ from $[0,+\infty)$ to $(0,1]$. $\lfloor\cdot\rfloor$ denotes the floor function, which assigns the style to one of $S$ categories. $PosEmb(\cdot)$ represents the positional encoding.

\begin{equation}
    \begin{aligned}
        & StyleEmb(\alpha) = PosEmb(\lfloor \bm{\alpha} S\rfloor)\\
        s.t.~&\bm{\alpha} = 1/(\alpha+1)
    \end{aligned}\label{eq:style-emb}
\end{equation}

\subsubsection{$\alpha$-DDIM Losses}

The loss function for the trajectory generation process consists of 3 components. The first component is the original loss function of DDIM, denoted as $L_{DDIM}$ in \cref{eq:l_1}. 
Secondly, $L_{sim}(\theta)$ constrains the gap between the human trajectory $\hat{X}$ and the machine trajectory ${p_a}$, as shown in equation \eqref{eq:l_2}.
The third component is the style loss $L_{style}$, responsible for providing constraints on $\alpha_t$, as illustrated in \cref{eq:l_3}. Where, $\alpha$ represents the theoretical complexity input into the model. 
The final loss $L$ is the sum of these components. $w_1, w_2$ and $w_3$ are the weights.

\begin{equation}
    L_{DDIM}:=\mathbb{E}_{\vec{X},\Sigma_\epsilon}\left[\Vert \Sigma_\epsilon-\Sigma(\sqrt{a}\vec{X}+\sqrt{1-a}\Sigma_\epsilon)\Vert_2\right]
    \label{eq:l_1}
\end{equation}

\begin{equation}
    L_{sim} := \Vert {p_a} - \hat{X}\Vert_2^2\label{eq:l_2}
\end{equation}

\begin{equation}
    L_{style} := \Vert \alpha - \frac{\sum_{i=1}^{m}\Vert p_a^{(i)}-p_a^{(i+1)}\Vert_2}{\Vert p_0-p_m\Vert_2} \Vert_2
    \label{eq:l_3}
\end{equation}

\begin{equation}
    L = w_1 L_{sim}(\theta) + w_2 L_{sim} + w_3\cdot L_{style}
\end{equation}

\subsection{Evaluation}

\subsubsection{Simulation Confidence Assessments}\label{sec:mtd-sca}

This subsection discusses solving RQ2 (White-box testing), which proposes to evaluate how well our model can generate the most human-like mouse trajectories. We devised two tests, one utilizing a distribution distance metric function and the other using a deep learning-based discriminator. The distribution distance metric measures the degree of similarity between the model-generated results and real human samples. The deep learning-based discriminator can find and amplify the small differences between the two trajectories and is more capable of determining whether the model simulates human operating habits in all aspects.

For ease of comparison, we will employ three sets of samples for measurement:
\begin{enumerate}
    \item \textbf{Human samples}: randomly sampled from collected human operation data.
    \item \textbf{DMTG samples}: multiple generation processes initialized based on each human sample, with one simulation result randomly selected from the outcomes.
    \item \textbf{Samples from other models}: an equal number of samples generated will be chosen from comparative generation models in the experiments.
\end{enumerate}

For distribution analysis, a variety of similarity and distance measurement methods will be employed, including Jensen–Shannon Divergence (JSD) \cite{jsd}, Earth Mover's Distance (EMD) \cite{Wasserstein,emd}, MSE, Root Mean Square Error (RMSE), Cosine Similarity (CosSim), and so on. These metrics will assess the disparities between model-generated mouse trajectories and human trajectories from multiple perspectives, such as curve density, mean, and distribution. Because our DMTG can control trajectory curves by complexity and length parameters, we will extract and scale the effective region during testing. Additionally, due to the bivariate Gaussian distribution nature of mouse trajectory data, which may complicate the computation of metrics like JSD, we will employ t-SNE \cite{tsne} to map high-dimensional curves to single-point vectors for computing.

For deep learning metrics, we designed two scenarios for robot recognition: unified and independent. The objective of the independent scenario design is to measure the degree of overlap between human and machine trajectories in data distribution. In this scenario, we employ randomly selected 7 sets of human and separate model samples as positive and negative samples, respectively, for training the discriminator to verify its recognition capability on the remaining 3 sets of samples. For the unified recognition scenario, we aim to demonstrate the effectiveness of our model in a global environment. Therefore, we construct a dataset with equal distribution comprising all human samples, DMTG samples, and samples from other models. The dataset is split into training and testing sets at a ratio of 7:3, to validate whether the discriminator can detect robot samples in the testing set.

\subsubsection{Commercial CAPTCHA Validations}\label{sec:mtd-ccv}

The primary bypassed commercial CAPTCHAs in this test are Akamai \cite{Akamai}, GeeTest \cite{Geetest}, and reCAPTCHA v3 \cite{google-recaptcha}. Akamai assigns a unique token to each session, collecting and uploading interaction data related to each web page operation to update the underlying confidence level of the session and determine whether to allow the request. Since the internal confidence level of Akamai cannot be known, only a signal indicating whether the test was passed will be returned. Therefore, experiments with Akamai will evaluate both the time taken to pass the test and the number of times the test is passed. The evaluation score is calculated using \cref{eq:akamai-s}, where $T$ is the theoretical time required for executing the trajectory, and $t_a$ is the execution time of the testing program. If no pass signal is issued after the trajectory execution is completed, the test is considered not to be passed. The final passing score will be the average score from multiple tests.

\begin{equation}
    S_{Akamai} = ReLU(\frac{T - t_a}{T})\label{eq:akamai-s}
\end{equation}

Geetest predicts and returns a confidence score of 1 to 6 for each attempt by the user to solve the CAPTCHA challenge, with higher scores being considered closer to human operation. Since we can access such scores, the average score over multiple attempts will be used as the evaluation.

 reCAPTCHA v3 uses IP credits and Google-related cookies for human-computer testing, and the product will return scores in the range of 0.1 to 0.7. In our testing, we found that mouse trajectory is not a significant factor for reCAPTCHA v3, and even human users received similar scores.

%% file: secs/exp.tex
\section{Experiments}\label{sec:exp}

This section will focus on experimental validation for RQ2 and RQ3, and will construct the experimental methodology following the process described in \cref{sec:mtd-sca,sec:mtd-ccv}. The comparative model for word experiments is as follows.

\begin{itemize}
    \item \textbf{GAN}: BeCAPTCHA-Mouse \cite{mt-becaptcha} utilizes GAN networks to simulate human mouse trajectories for detecting robot operations.
    \item \textbf{SapiAgent}: SapiAgent \cite{mt-SapiAgent} employs the SapiMouse \cite{SapiMouse} dataset and a deep encoder architecture for CAPTCHA deceptive bypass.
    \item \textbf{Bezier}: Bezier \cite{bezier} is a method based on generating smooth curves using multiple anchor points.
    \item \textbf{Linear}: This method, discussed in \cref{sec:mtd-dmtg}, simply employs a control approach using $N$ equidistant points along the $\vec{X}$ direction.
    \item \textbf{Ghost}: Ghost-cursor \cite{ghost} dynamically adjusts the displacement velocity between mouse trajectory curves and operating DOM elements using Fitts' Law to create realistic behavior. This method represents a relatively successful engineering case.
\end{itemize}

The deep learning trajectory discriminators in \cref{sec:mtd-sca} will be implemented using Decision Tree (DT) \cite{dt}, Random Forest (RF) \cite{rf}, XGBoost (XGB) \cite{XGBoost}, Gradient Boosting (GB) \cite{graboost}, MLP \cite{MLP}, CNN \cite{CNN}, RNN \cite{RNN}, LSTM \cite{LSTM}, BiLSTM \cite{BiLSTM}, and TCN \cite{TCN}. Each discriminator will carry out experiments on all comparison models and DMTGs separately following the methods involved in \cref{sec:mtd-sca}. All the deep learning models are built using PyTorch and trained on a Nvidia A100 GPU.

\begin{figure*}
    \centering
    \subfigure[$\bm{\alpha}=0.1$]{
        \includegraphics[width=.15\linewidth]{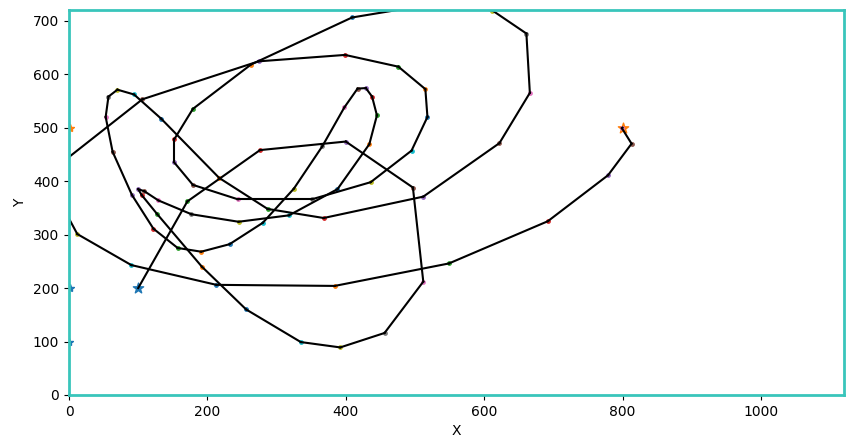}
        \includegraphics[width=.15\linewidth]{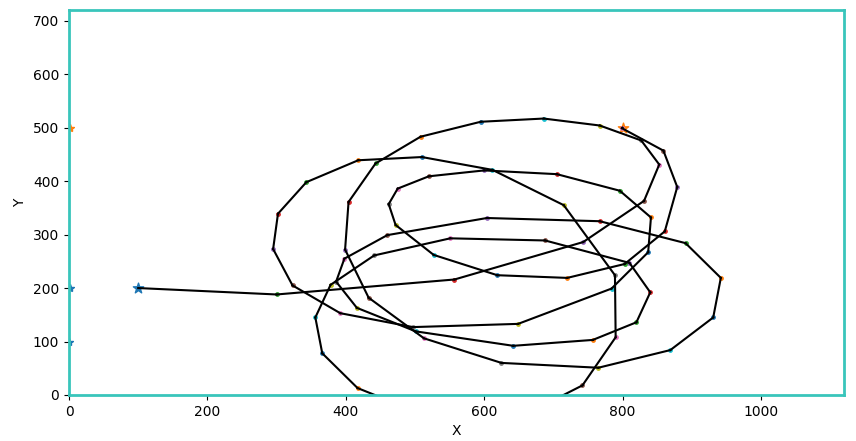} 
        \includegraphics[width=.15\linewidth]{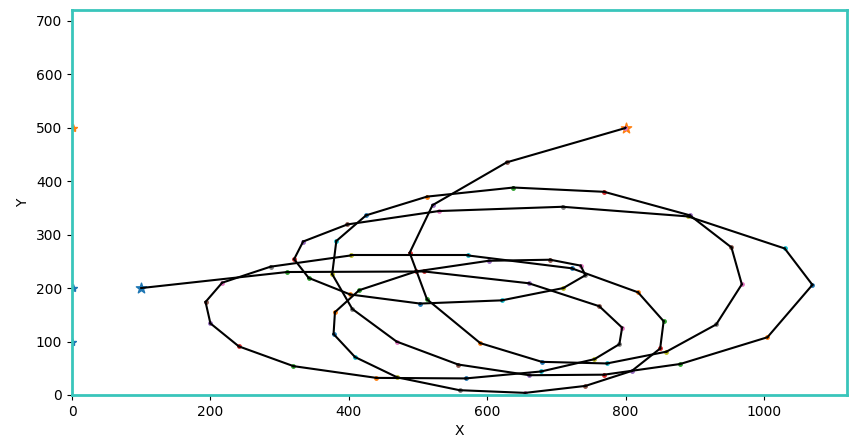}
        \label{fig:alpha-0.1-curves}
    }
    \subfigure[$\bm{\alpha}=0.2$]{
        \includegraphics[width=.15\linewidth]{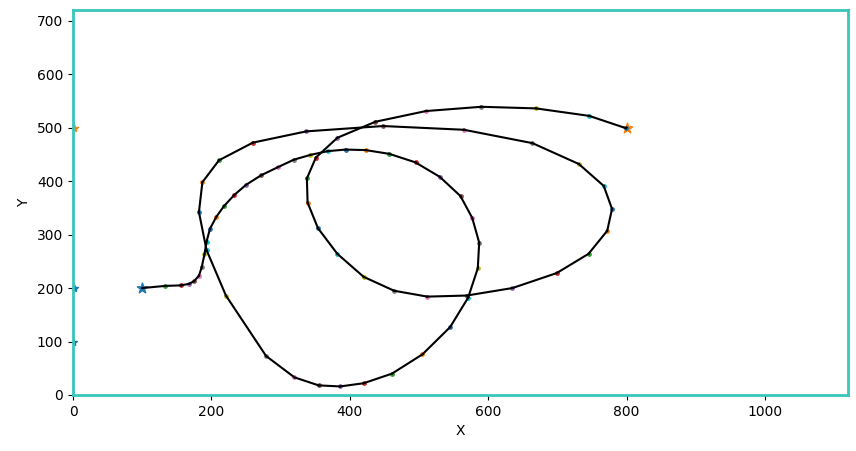}
        \includegraphics[width=.15\linewidth]{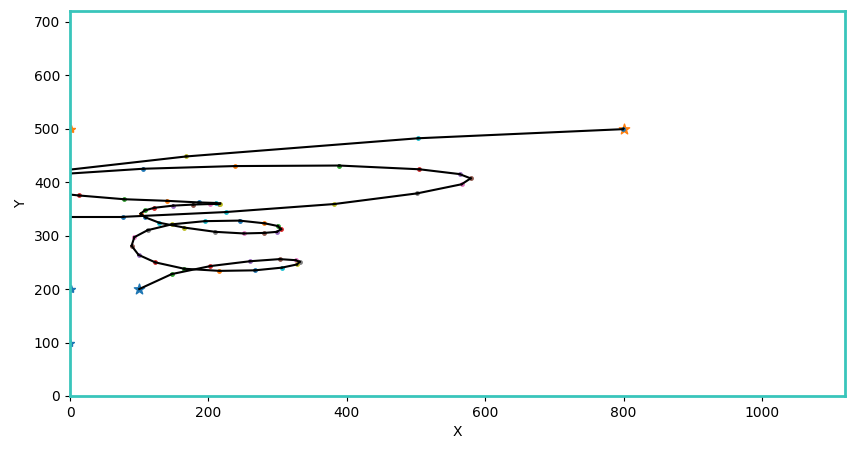}
        \includegraphics[width=.15\linewidth]{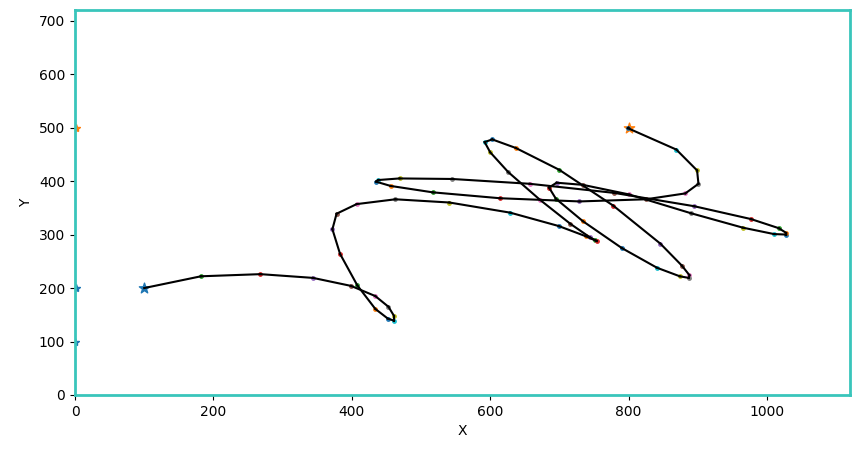}
        \label{fig:alpha-0.2-curves}
    }
    \subfigure[$\bm{\alpha}=0.3$]{
        \includegraphics[width=.15\linewidth]{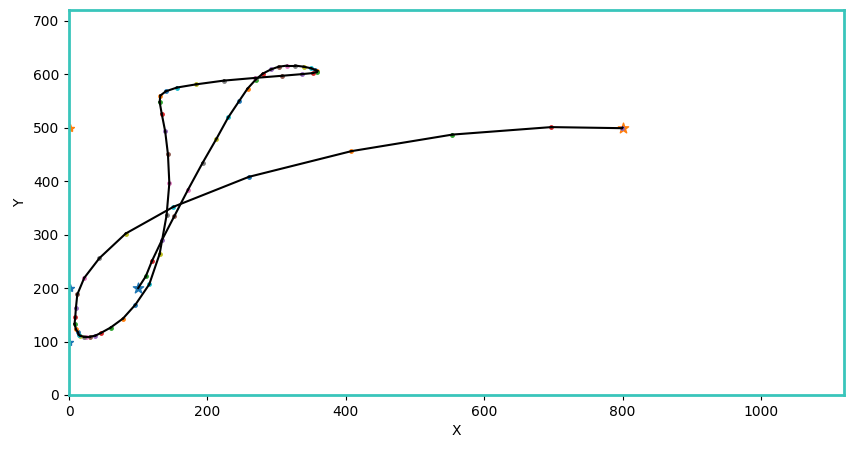}
        \includegraphics[width=.15\linewidth]{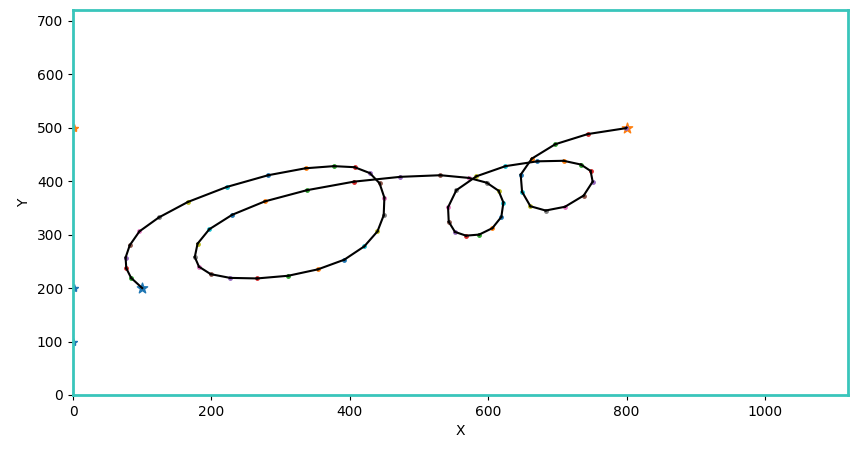}
        \includegraphics[width=.15\linewidth]{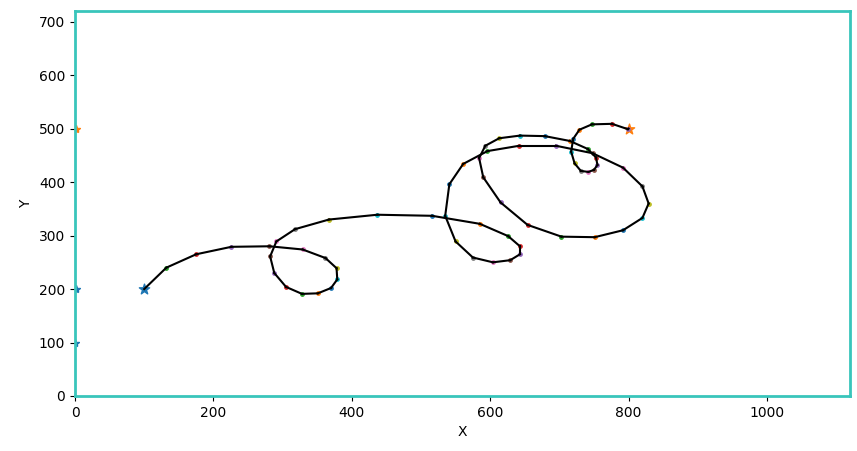}
        \label{fig:alpha-0.3-curves}
    }
    \subfigure[$\bm{\alpha}=0.4$]{
        \includegraphics[width=.15\linewidth]{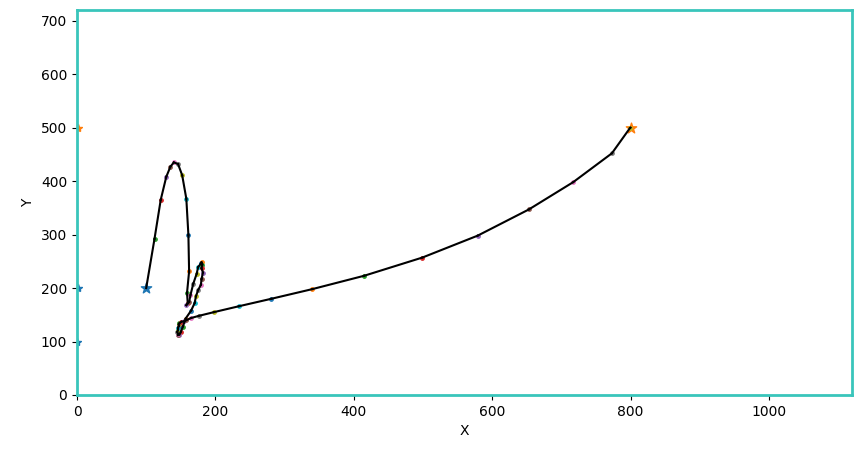}
        \includegraphics[width=.15\linewidth]{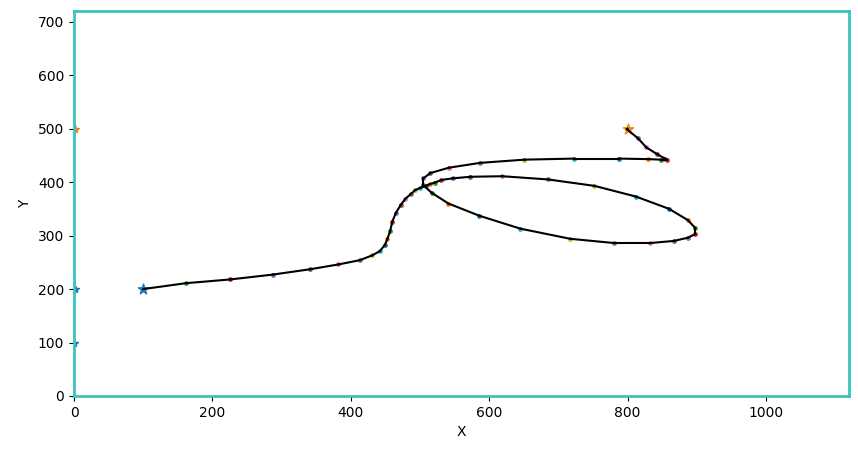}
        \includegraphics[width=.15\linewidth]{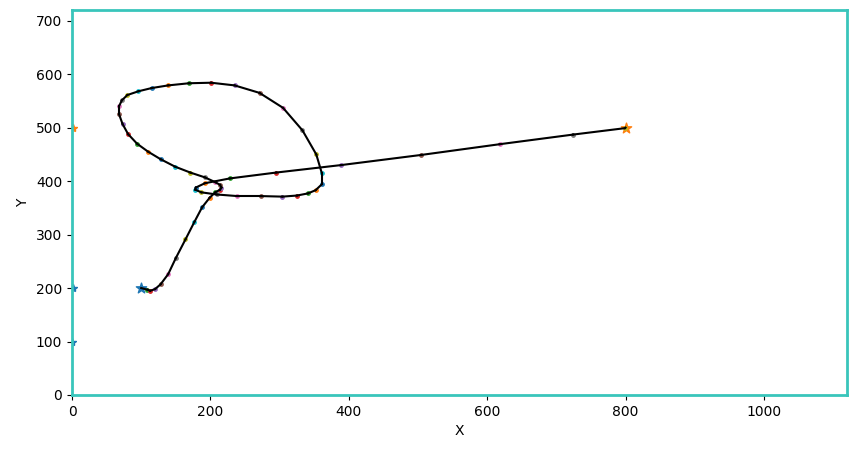}
        \label{fig:alpha-0.4-curves}
    }
    \subfigure[$\bm{\alpha}=0.5$]{
        \includegraphics[width=.15\linewidth]{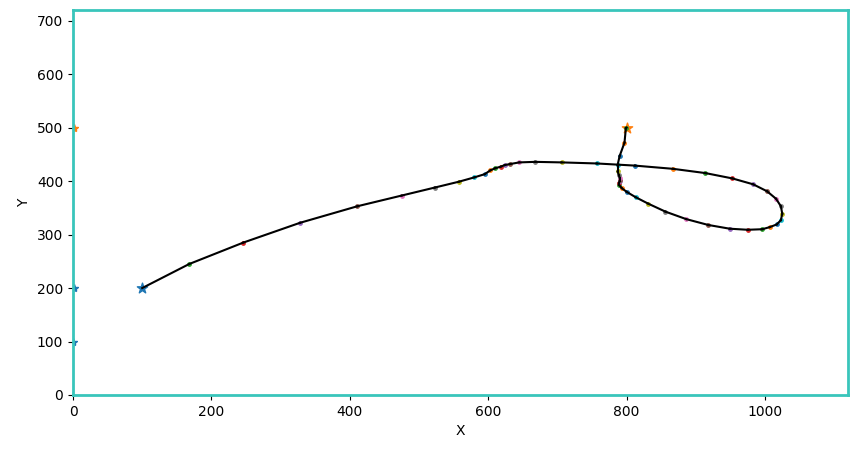}
        \includegraphics[width=.15\linewidth]{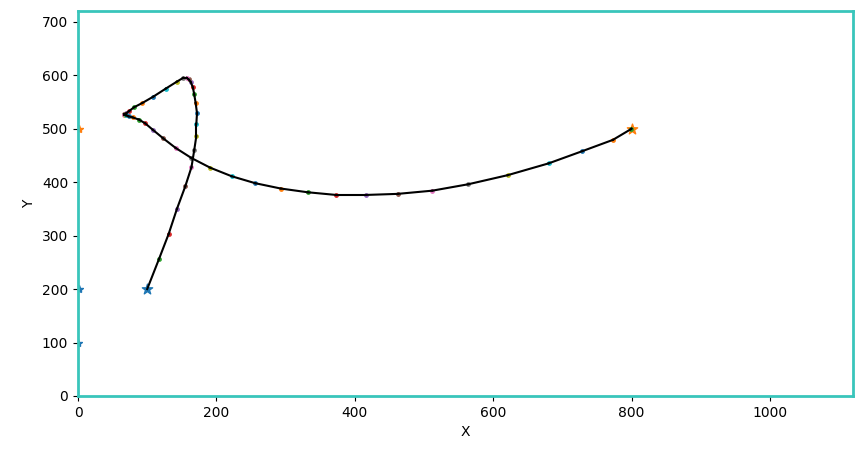}
        \includegraphics[width=.15\linewidth]{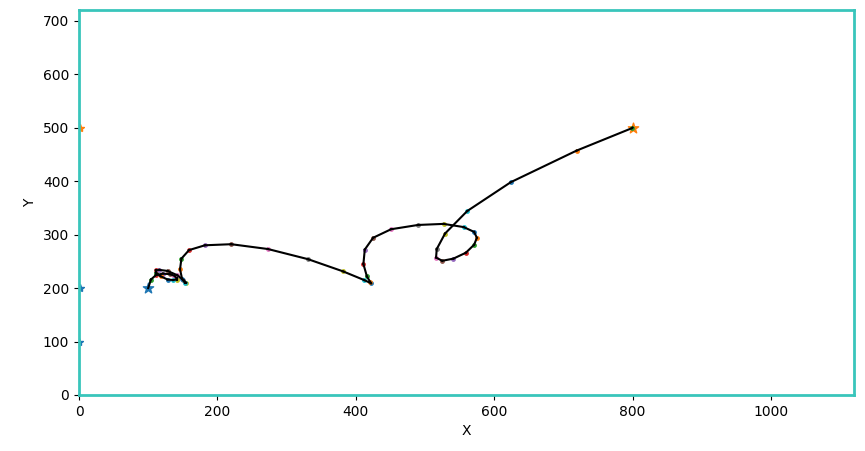}
        \label{fig:alpha-0.5-curves}
    }
    \subfigure[$\bm{\alpha}=0.6$]{
        \includegraphics[width=.15\linewidth]{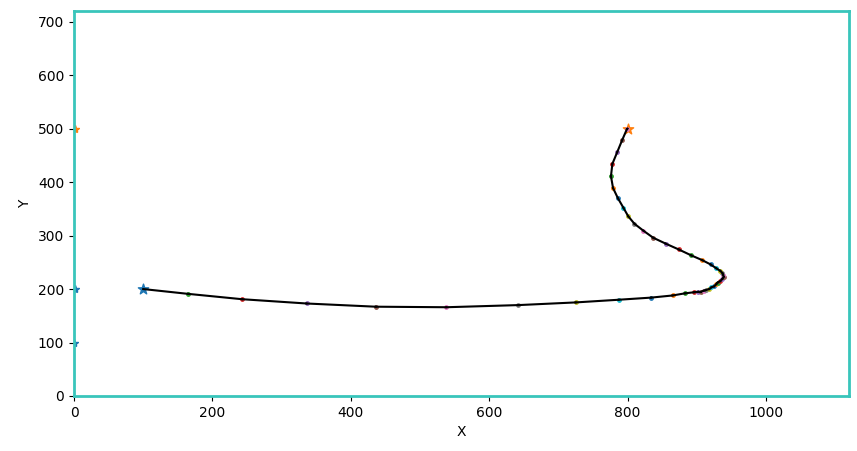}
        \includegraphics[width=.15\linewidth]{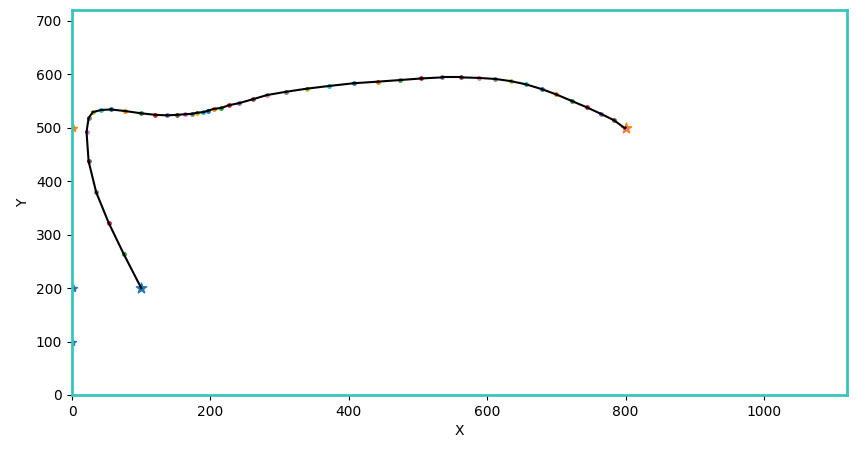}
        \includegraphics[width=.15\linewidth]{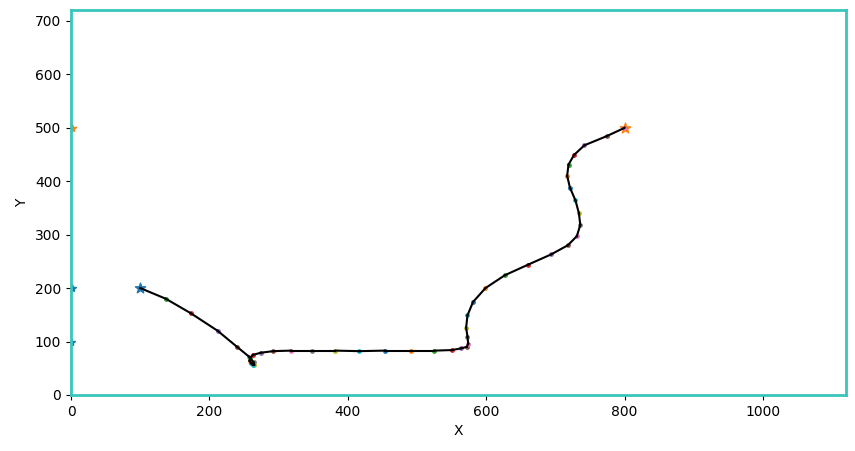}
        \label{fig:alpha-0.6-curves}
    }
    \subfigure[$\bm{\alpha}=0.7$]{
        \includegraphics[width=.15\linewidth]{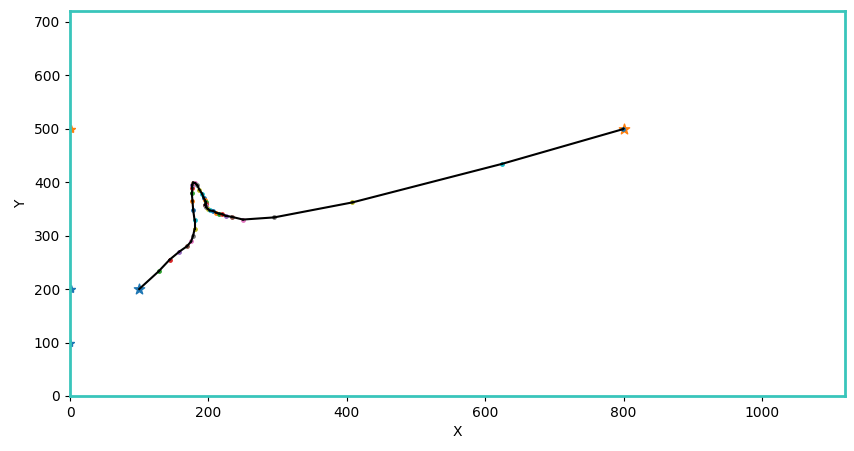}
        \includegraphics[width=.15\linewidth]{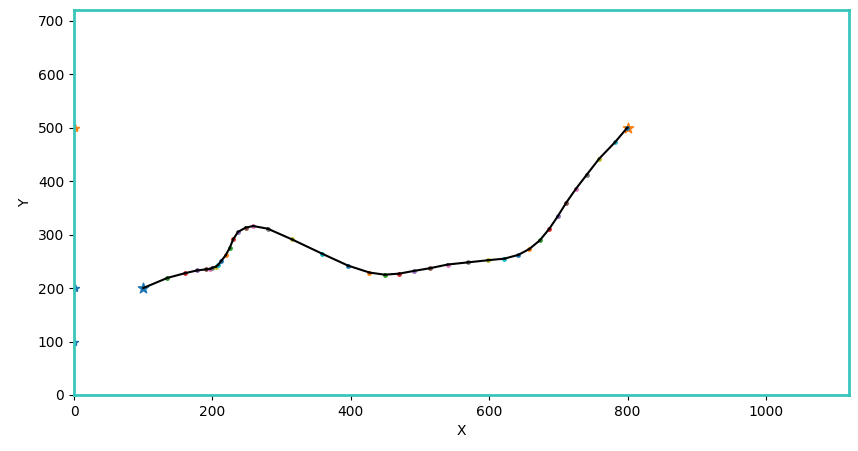}
        \includegraphics[width=.15\linewidth]{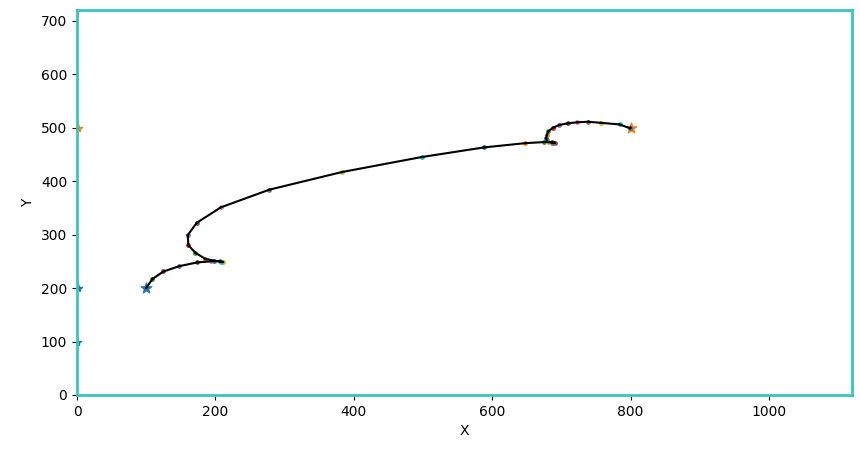}
        \label{fig:alpha-0.7-curves}
    }
    \subfigure[$\bm{\alpha}=0.8$]{
        \includegraphics[width=.15\linewidth]{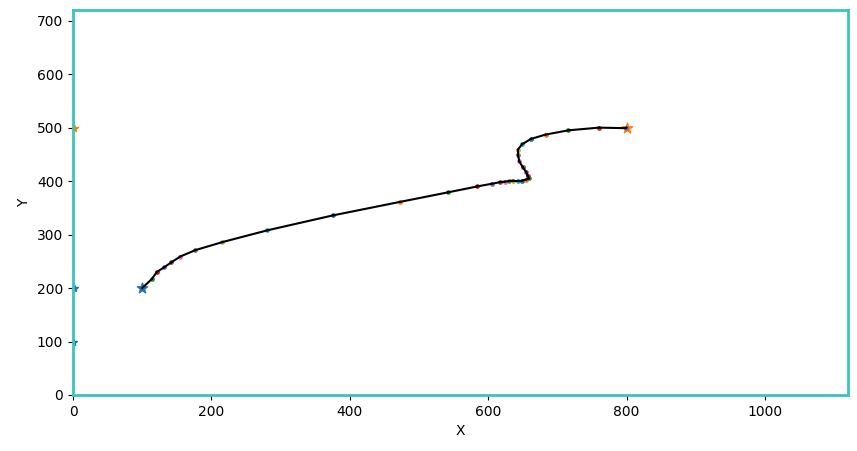}
        \includegraphics[width=.15\linewidth]{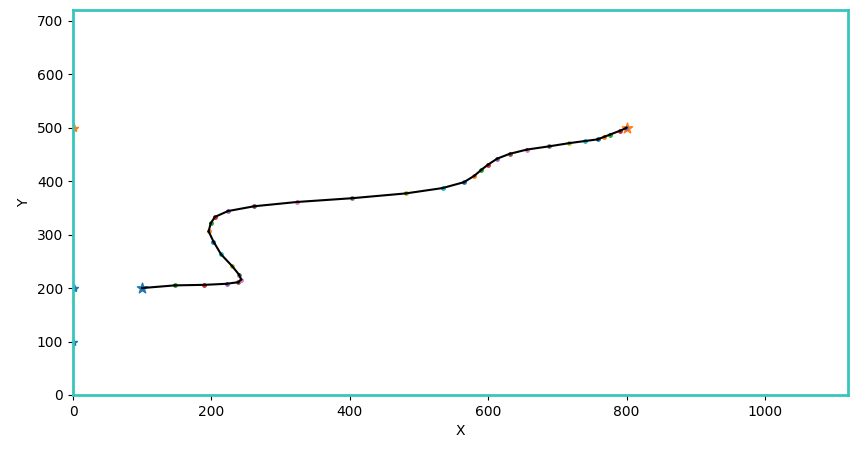}
        \includegraphics[width=.15\linewidth]{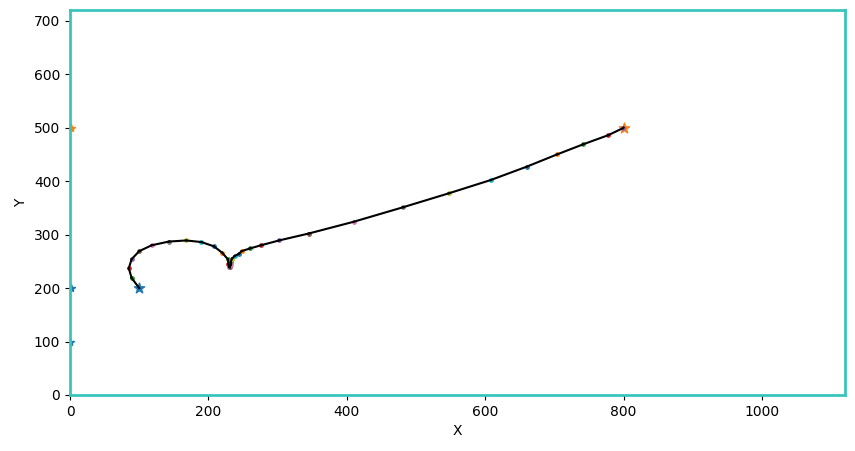}
        \label{fig:alpha-0.8-curves}
    }
    \subfigure[$\bm{\alpha}=0.9$]{
        \includegraphics[width=.15\linewidth]{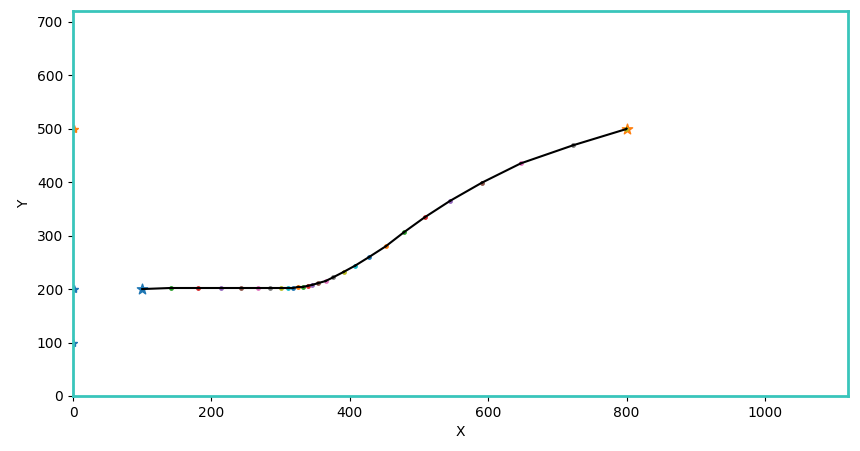}
        \includegraphics[width=.15\linewidth]{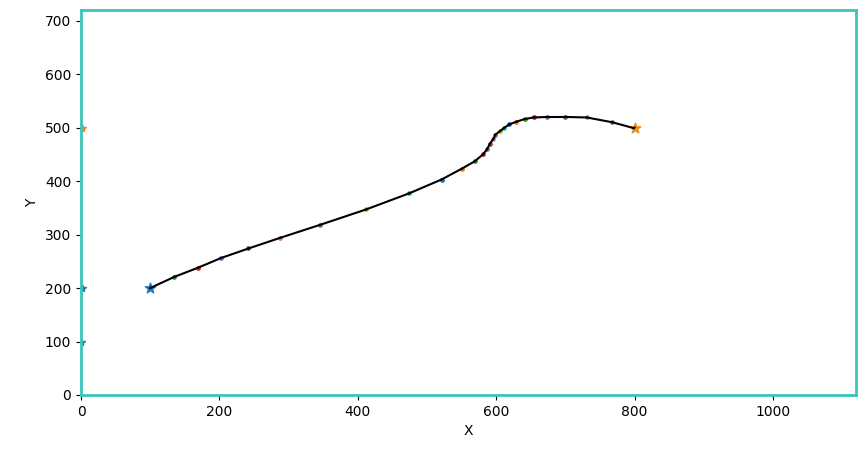}
        \includegraphics[width=.15\linewidth]{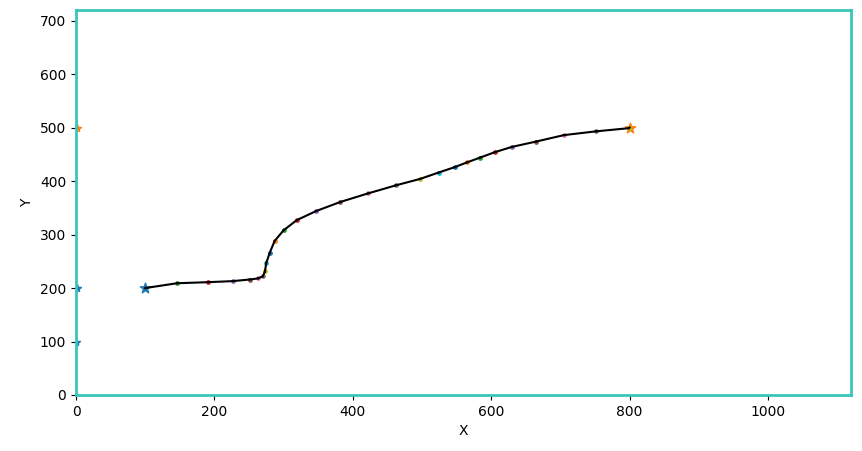}
        \label{fig:alpha-0.9-curves}
    }
    \subfigure[$\bm{\alpha}=1.0$]{
        \includegraphics[width=.15\linewidth]{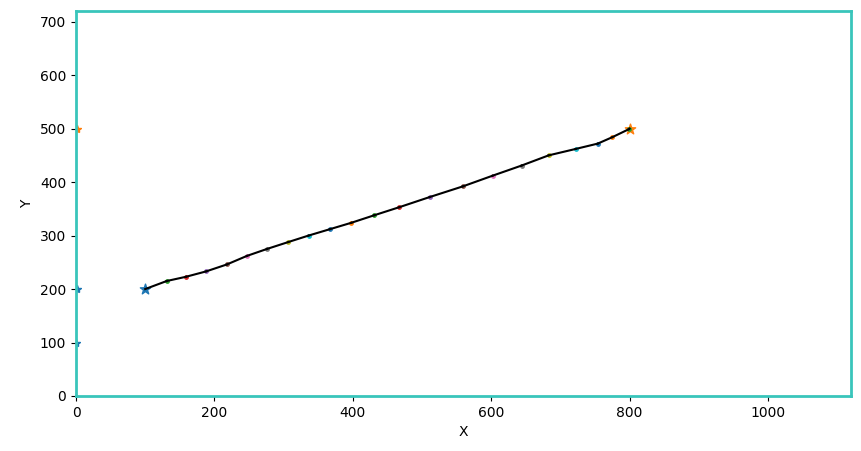}
        \includegraphics[width=.15\linewidth]{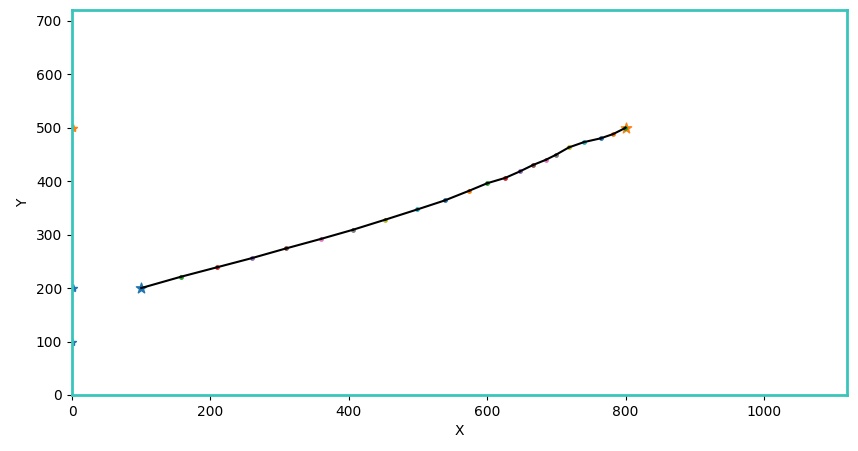}
        \includegraphics[width=.15\linewidth]{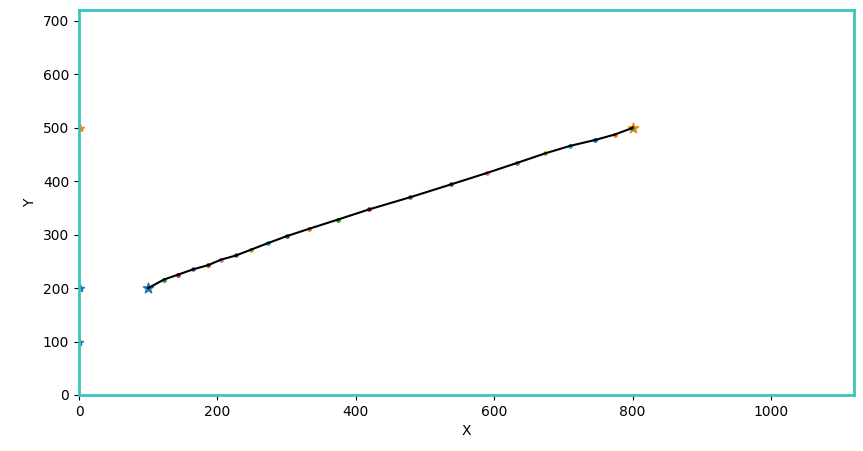}
        \label{fig:alpha-1.0-curves}
    }
    \caption{Examples of DMTG Curves under Different $\bm{\alpha}$ Values in \cref{eq:style-emb}}
    \label{fig:alpha-curves}
\end{figure*}

\begin{table*}
    \centering
    \caption{Differences in Statistics with Different $\bm{\alpha}$s}
    \begin{tabular}{cccccc|cccc}
        \toprule
        \multirow{2}{*}{Metric} & \multicolumn{9}{c}{$\bm{\alpha}$ in \cref{eq:style-emb}} \\
        \cline{2-10}
        & 0.0-0.2 & 0.2-0.4 & 0.4-0.6 & 0.6-0.8 & 0.8-1.0 & 0.0-0.1 & 0.1-0.2 & 0.2-0.3 & 0.3-0.4 \\
        \midrule
        JSD & 0.4520 & \uline{0.4293} & 0.4560 & 0.5194 & 0.5528 & 0.5659 & 0.4845 & \textbf{0.3911} & 0.5131 \\
        EMD & 6.9235e-4 & \uline{4.8889e-4} & 5.1111e-4 & 6.4444e-4 & 1.0444e-3 & 6.1111e-4 & 9.0000e-4 & \textbf{3.1111e-4} & 1.2778e-3 \\
        MSE & 1.1164e+3 & \textbf{9.9901e+2} & 1.0660e+3 & 1.0254e+3 & 1.1488e+3 & 4.3043e+3 & 1.7324e+3 & \uline{1.3168e+3} & 1.49763+3 \\
        RMSE & 3.3412e+1 & \textbf{3.1607e+1} & 3.2650e+1 & 3.2021e+1 & 3.3893e+1 & 6.5638e+1 & 4.1622e+1 & \uline{3.6288e+1} & 3.8699e+1 \\
        CosSim & -2.8197e-2 & \textbf{5.0583e-3} & -9.8744e-2 & -5.4786e-2 & -2.0567e-2 & \uline{-2.4123e-2} & -3.4053e-2 & -2.4335e-2 & -8.8296e-2 \\
        \bottomrule
    \end{tabular}
    \label{tab:alpha-metrics}
\end{table*}

Before formal testing, we wish to determine which complexity configuration can more easily yield human-like curves. Therefore, we analyzed the influence of different values of $\bm{\alpha}$ in \cref{eq:style-emb} based on human datasets. The results are shown in \cref{fig:alpha-curves} and \cref{tab:alpha-metrics}. According to the results, we found that when $\bm{\alpha}\leq 0.2$, the generated curves exhibit excessively high randomness. While when $\bm{\alpha}\geq 0.8$, the generated curves tend to approximate a straight line. The optimal curves correspond to $\bm{\alpha}$ values between 0.2 and 0.3. Hence, we will use random values ranging from 0.3 to 0.8 as $\bm{\alpha}$ for the subsequent experiments.

\subsection{RQ2: Simulation Confidence Assessments}

\begin{table}
    \centering
    \caption{Prediction Indicators of Independent Discriminators for Distinguishing Humans from Machines}
    \begin{tabular}{ccccccc}
        \toprule
        \multirow{2}{*}{Detector} & \multicolumn{6}{c}{Generator} \\
        \cline{2-7}
         & DMTG & GAN & SapiAgent & Bezier & Ghost & Linear \\
        \midrule
        \multicolumn{7}{c}{Accuracy} \\
        \midrule
        DT & \textbf{0.8759} & 0.9993 & 0.9961 & 0.9776 & \uline{0.9310} & 0.9810 \\
        RF & \textbf{0.9149} & 0.9994 & 0.9993 & 0.9845 & \uline{0.9558} & 0.9858 \\
        GB & \textbf{0.8912} & 0.9993 & 0.9973 & 0.9624 & \uline{0.9290} & 0.9609 \\
        XGB & \textbf{0.9194} & 0.9996 & 0.9995 & 0.9874 & \uline{0.9608} & 0.9869 \\
        MLP & \textbf{0.8677} & 0.9989 & 0.9932 & 0.9677 & \uline{0.9275} & 0.9861 \\
        CNN & \textbf{0.8920} & 0.9995 & 0.9912 & 0.9869 & \uline{0.9563} & 0.9971 \\
        RNN & \textbf{0.8804} & 0.9976 & 0.9905 & 0.9677 & \uline{0.8900} & 0.9050 \\
        LSTM & \textbf{0.8709} & 0.9982 & 0.9918 & 0.9729 & \uline{0.9259} & 0.9923 \\
        BiLSTM & \uline{0.9011} & 0.9940 & \textbf{0.8880} & 0.9632 & 0.9646 & 0.9980 \\
        TCN & \textbf{0.9051} & 0.9992 & 0.9957 & 0.9852 & \uline{0.9566} & 0.9881 \\
        \midrule
        \multicolumn{7}{c}{Macro Precision} \\
        \midrule
        MLP & \textbf{0.8836} & 0.9989 & 0.9932 & 0.9691 & \uline{0.9280} & 0.9865 \\
        CNN & \textbf{0.8944} & 0.9995 & 0.9912 & 0.9870 & \uline{0.9577} & 0.9971 \\
        RNN & \textbf{0.8977} & 0.9976 & 0.9905 & 0.9678 & \uline{0.9095} & 0.9118 \\
        LSTM & \textbf{0.8846} & 0.9982 & 0.9918 & 0.9738 & \uline{0.9279} & 0.9924 \\
        BiLSTM & \textbf{0.9018} & 0.9940 & \uline{0.9083} & 0.9644 & 0.9648 & 0.9980 \\
        TCN & \textbf{0.9095} & 0.9992 & 0.9957 & 0.9856 & \uline{0.9567} & 0.9883 \\
        \midrule
        \multicolumn{7}{c}{Macro Recall} \\
        \midrule
        MLP & \textbf{0.8677} & 0.9989 & 0.9932 & 0.9677 & \uline{0.9275} & 0.9861 \\
        CNN & \textbf{0.8920} & 0.9995 & 0.9912 & 0.9869 & \uline{0.9563} & 0.9971 \\
        RNN & \textbf{0.8804} & 0.9976 & 0.9905 & 0.9677 & \uline{0.8900} & 0.9050 \\
        LSTM & \textbf{0.8709} & 0.9982 & 0.9918 & 0.9729 & \uline{0.9259} & 0.9923 \\
        BiLSTM & \uline{0.9011} & 0.9940 & \textbf{0.8880} & 0.9632 & 0.9646 & 0.9980 \\
        TCN & \textbf{0.9051} & 0.9992 & 0.9957 & 0.9852 & \uline{0.9566} & 0.9881 \\
        \midrule
        \multicolumn{7}{c}{Macro F1} \\
        \midrule
        MLP & \textbf{0.8663} & 0.9989 & 0.9932 & 0.9677 & \uline{0.9274} & 0.9861 \\
        CNN & \textbf{0.8918} & 0.9995 & 0.9912 & 0.9869 & \uline{0.9563} & 0.9971 \\
        RNN & \textbf{0.8791} & 0.9976 & 0.9905 & 0.9677 & \uline{0.8886} & 0.9046 \\
        LSTM & \textbf{0.8697} & 0.9982 & 0.9918 & 0.9729 & \uline{0.9258} & 0.9923 \\
        BiLSTM & \uline{0.9010} & 0.9940 & \textbf{0.8866} & 0.9632 & 0.9646 & 0.9980 \\
        TCN & \textbf{0.9049} & 0.9992 & 0.9957 & 0.9852 & \uline{0.9566} & 0.9881 \\
        \bottomrule
    \end{tabular}
    \label{tab:indep-discrim}
\end{table}

\begin{table}
    \centering
    \caption{Prediction Indicators of Unified Discriminators for Distinguishing Humans from Machines}
    \begin{tabular}{ccccccc}
        \toprule
        \multirow{2}{*}{Detector} & \multicolumn{6}{c}{Generator} \\
        \cline{2-7}
        & DMTG & GAN & SapiAgent & Bezier & Ghost & Linear \\
        \midrule
        \multicolumn{7}{c}{Accuracy} \\
        \midrule
        DT & \textbf{0.8210} & 0.9586 & 0.9548 & 0.9401 & \uline{0.8993} & 0.9416 \\
        RF & \textbf{0.8655} & 0.9747 & 0.9736 & 0.9726 & \uline{0.9385} & 0.9724 \\
        GB & \textbf{0.8276} & 0.9500 & 0.9331 & 0.9359 & \uline{0.8922} & 0.9333 \\
        XGB & \textbf{0.8763} & 0.9730 & 0.9709 & 0.9745 & \uline{0.9422} & 0.9704 \\
        MLP & \textbf{0.7646} & 0.9481 & 0.9376 & 0.9369 & \uline{0.8370} & 0.9481 \\
        CNN & \textbf{0.7921} & 0.9658 & 0.9599 & 0.9544 & \uline{0.9262} & 0.9567 \\
        RNN & \textbf{0.7257} & 0.9665 & 0.8358 & 0.9323 & \uline{0.8461} & 0.9613 \\
        LSTM & \textbf{0.7765} & 0.9390 & 0.9376 & 0.9397 & \uline{0.9177} & 0.9554 \\
        BiLSTM & \textbf{0.8145} & 0.9740 & 0.9485 & 0.9594 & \uline{0.9405} & 0.9640 \\
        TCN & \textbf{0.8346} & 0.9636 & 0.9640 & 0.9502 & \uline{0.8900} & 0.9549 \\
        \midrule
        \multicolumn{7}{c}{Macro Precision} \\
        \midrule
        MLP & \textbf{0.7852} & 0.9523 & 0.9399 & 0.9390 & \uline{0.8417} & 0.9521 \\
        CNN & \textbf{0.8169} & 0.9677 & 0.9623 & 0.9575 & \uline{0.9255} & 0.9600 \\
        RNN & \textbf{0.7991} & 0.9683 & 0.8475 & 0.9355 & \uline{0.8707} & 0.9617 \\
        LSTM & \textbf{0.8012} & 0.9447 & 0.9431 & 0.9403 & \uline{0.9177} & 0.9577 \\
        BiLSTM & \textbf{0.8317} & 0.9752 & 0.9494 & 0.9608 & \uline{0.9407} & 0.9661 \\
        TCN & \textbf{0.8453} & 0.9659 & 0.9659 & 0.9529 & \uline{0.8906} & 0.9582 \\
        \midrule
        \multicolumn{7}{c}{Macro Recall} \\
        \midrule
        MLP & \textbf{0.7646} & 0.9481 & 0.9376 & 0.9369 & \uline{0.8370} & 0.9481 \\
        CNN & \textbf{0.7921} & 0.9658 & 0.9599 & 0.9544 & \uline{0.9262} & 0.9567 \\
        RNN & \textbf{0.7257} & 0.9665 & 0.8358 & 0.9323 & \uline{0.8461} & 0.9613 \\
        LSTM & \textbf{0.7765} & 0.9390 & 0.9376 & 0.9397 & \uline{0.9177} & 0.9554 \\
        BiLSTM & \textbf{0.8145} & 0.9740 & 0.9485 & 0.9594 & \uline{0.9405} & 0.9640 \\
        TCN & \textbf{0.8346} & 0.9636 & 0.9640 & 0.9502 & \uline{0.8900} & 0.9549 \\
        \midrule
        \multicolumn{7}{c}{Macro F1} \\
        \midrule
        MLP & \textbf{0.7603} & 0.9480 & 0.9376 & 0.9369 & \uline{0.8365} & 0.9480 \\
        CNN & \textbf{0.7879} & 0.9658 & 0.9598 & 0.9543 & \uline{0.9252} & 0.9566 \\
        RNN & \textbf{0.7078} & 0.9665 & 0.8344 & 0.9321 & \uline{0.8435} & 0.9613 \\
        LSTM & \textbf{0.7718} & 0.9388 & 0.9364 & 0.9397 & \uline{0.9177} & 0.9553 \\
        BiLSTM & \textbf{0.8121} & 0.9740 & 0.9485 & 0.9593 & \uline{0.9405} & 0.9640 \\
        TCN & \textbf{0.8333} & 0.9636 & 0.9639 & 0.9502 & \uline{0.8900} & 0.9548 \\
        \bottomrule
    \end{tabular}
    \label{tab:unify-discrim}
\end{table}

\cref{tab:indep-discrim,tab:unify-discrim} respectively present the prediction accuracy and other metrics of human and machine trajectories using unified and independent discriminators. For each test, every model generates a sample size of 1 million for random selection. Higher accuracy indicates that the model curves and true mouse trajectories are more easily distinguishable by the discriminator. Therefore, lower scores in the table indicate that the generated curves by the models have smaller differences in features compared to human operations. The following conclusions can be drawn: 

\begin{enumerate}
    \item \textbf{DMTG performs the best among models}. Various discriminators show the lowest prediction accuracy for our DMTG model, approximately 10\% lower than other confusion methods. This indicates that our trajectories are most prone to confusion with human actions, thereby misleading the discriminator to make incorrect classification predictions. Although Ghost can also provide trajectories that are easily misclassified, its confusion strength is much weaker than DMTG.
    \item \textbf{DMTG does not introduce additional errors}. Although visually and intuitively, GAN and SapiAgent can generate highly complex and sufficiently simulated curves, their trajectories are significantly more likely to be discovered than Bezier and Linear. This indirectly reflects that these curves cannot reliably reflect the distribution of real mouse trajectories. Instead, they are more easily discovered due to the introduction of more obvious trajectory noise.
    \item \textbf{DMTG is more suitable for online detection scenarios}. When using independent detectors, the discriminator's recognition accuracy for DMTG is around 90\%; however, when using a unified detector, the accuracy drops to around 80\%. Under the same conditions, Ghost's accuracy drops from around 92\% to 90\%, and the average accuracy of other models drops from 99\% to 95\%, with reductions of less than 10\% on average. This indicates that if online detectors use multiple judgment methods for black-box testing, similar to the working principle of a unified discriminator, DMTG will perform better than the current white-box testing.
\end{enumerate}

\begin{table}
    \centering
    \caption{Adaptation and Retention of Parameter Controls for Each Generating Model Based on Deep Learnings}
    \begin{tabular}{cccc}
        \toprule
        \multirow{2}{*}{Model} & \multicolumn{3}{c}{Control Parameter} \\
        \cline{2-4}
        & Effective Length & Start \& End Points & Complexity \\
        \midrule
        DMTG & \textbf{0} & \textbf{8.2784e-10} & \textbf{0.0179} \\
        SapiAgent & 102 & 0.3383 & $\times$ \\
        GAN & \textbf{0} & $\times$ & $\times$ \\
        \bottomrule
    \end{tabular}
    \label{tab:parameter-control}
\end{table}

\cref{tab:parameter-control} illustrates the acceptance levels of different deep learning models towards various curve control parameters. The symbol ``$\times$" indicates that the method cannot accept control of the corresponding parameter, while numbers denote the MSE disparity between expected input and generated results. Regarding effective length control, although all models can accept the Mask matrix to constrain the number of effective nodes, SapiAgent exhibits slight jitter at locations where it should remain locked during iterations, leading to its average length exceeding the restriction. This issue also contributes to errors in the starting point control of SapiAgent. In summary, our method can accept all parameter constraints related to curve control, and it outperforms others in maintaining parameter effectiveness.

\cref{fig:tsne-curves} illustrates the disparities between mouse trajectories generated by three deep learning models and Ghost, compared to human data. The blue area represents the distribution of human data, while the red dots depict the distribution of model results. From these results, several observations can be made:

\begin{enumerate}
    \item Compared to SapiAgent and GAN, the data generated by DMTG and Ghost are closer to the distribution of human data. 
    \item \textbf{The distribution center of DMTG is closer to the mean of the human distribution.} From the graph, it is evident that although Ghost and our method both overlap with human data, Ghost's data is noticeably concentrated at one end of the human distribution area. This results in the overlapping area appearing as one end red and the other end blue. Conversely, the distribution of our model's results exhibits a higher degree of blending with the human distribution, covering all regions of the human distribution. This is reflected in the graph as DMTG's distribution ``enclosing" the human distribution, displaying an outer layer of red and an inner layer of blue.
\end{enumerate}

\begin{table}
    \centering
    \caption{Differences in Statistics between Mouse Trajectories of Humans and Various Models on Data Distribution}
    \begin{tabular}{ccccc}
        \toprule
        \multirow{2}{*}{Metric} & \multicolumn{4}{c}{Generator} \\
        \cline{2-5}
          & DMTG & GAN & Ghost & SapiAgent \\
        \midrule
        JSD & \textbf{0.3025} & 0.5865 & \uline{0.3565} & 0.6325 \\
        EMD & \textbf{1.5170e-4} & 7.9644e-4 & \uline{2.5896e-4} & 1.4080e-3 \\
        MSE & \textbf{5.5919e+3} & 6.5803e+3 & \uline{6.0058e+3} & 6.6803e+3 \\
        RMSE & \textbf{7.4779e+1} & 8.1191e+1 & \uline{7.7497e+1} & 8.1733e+1 \\
        CosSim & \textbf{-0.0130} & -0.1829 & \uline{-0.0130} & -0.1356 \\
        \bottomrule
    \end{tabular}
    \label{tab:math-metrics}
\end{table}

\cref{tab:math-metrics} displays the discrepancies between the results generated by all models and human data in terms of data statistics. It can be observed that our model consistently ranks favorably across various evaluation metrics, indicating that trajectories generated by DMTG closely resemble those of humans. In addition, DMTG and Ghost exhibit significant advantages over the other two similarity metrics related to data distribution, such as JSD and EMD. Although there is little difference among the four models in metrics related to node distance and density such as MSE and RMSE, our model still exhibits the smallest errors.

\subsection{RQ3: Commercial CAPTCHA Validations}
\textbf{To further verify the effectiveness of DMTG in real-world applications, we tested it on commercial CAPTCHA systems such as GeeTest and Akamai. These evaluations provided valuable insights into the performance of our model. However, due to the copyright and proprietary restrictions involved with these services, we cannot disclose detailed methods and results in this paper. We understand the importance of transparency and plan to provide more detailed analysis in future work when possible.}

\begin{figure*}
    \centering
    \subfigure[Human]{
        \includegraphics[width=.4\linewidth]{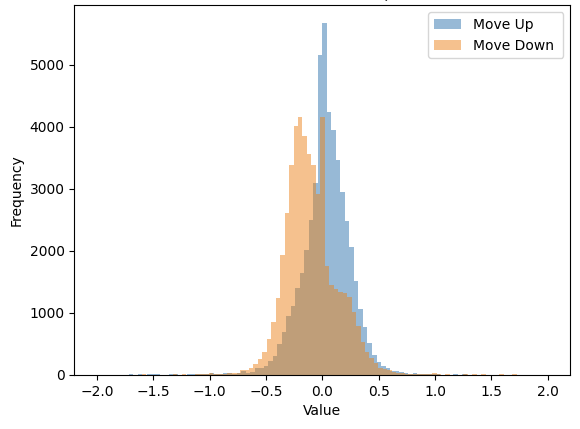}
        \label{fig:human-mouse-updown}
    }
    \subfigure[DMTG]{
        \includegraphics[width=.4\linewidth]{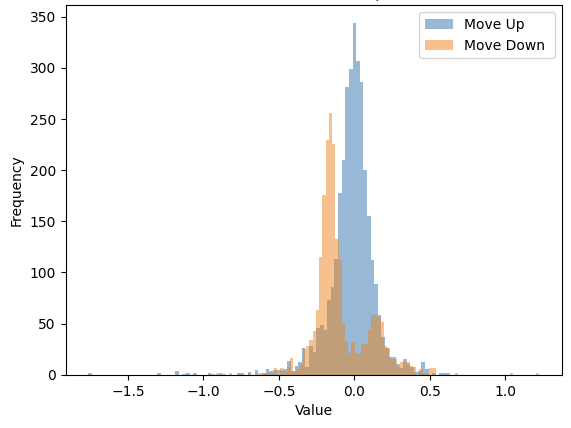}
        \label{fig:DMTG-mouse-updown}
    }
    \subfigure[SapiAgent]{
        \includegraphics[width=.4\linewidth]{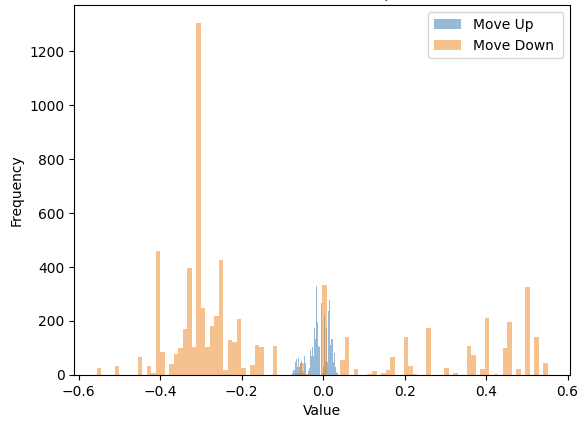}
        \label{fig:sapi-mouse-updown}
    }
    \subfigure[GAN]{
        \includegraphics[width=.4\linewidth]{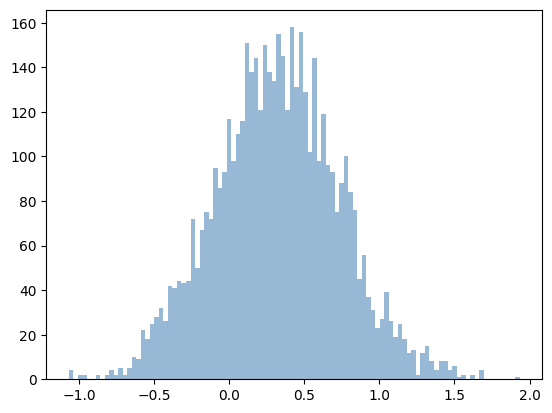}
        \label{fig:gan-mouse-updown}
    }
    \caption{Difference in Acceleration between Upward and Downward Mouse Movements}
    \label{fig:mouse-updown}
\end{figure*}

To understand why our model can generate more realistic trajectories, we conducted a statistical analysis of the acceleration distribution of mouse movements. The comparative results among models are depicted in \cref{fig:mouse-updown}. Except for GAN, all other models can distinguish different movement directions. It is certain that, for humans, the acceleration during upward mouse movements differs from that during downward activation. This discrepancy arises because the force applied to the mouse varies when pushing forward versus pulling backward. Among the models capable of generating random trajectories, only our model can effectively capture this characteristic. This fundamental difference is the reason behind the superior performance of DMTG compared to other models.

\subsection{Answers to RQs}

For the three research questions we posed, we provide the following answers:
\begin{enumerate}
    \item \textbf{Answers to RQ1}: We introduced a mouse trajectory generation method called DMTG, which can control the complexity of curves. It can closely mimic human actions, including acceleration during movement, trajectory uncertainty, and consistency with targets.
    \item \textbf{Answers to RQ2}: We proposed a white-box evaluation method combining mathematical statistics and deep learning to measure the model's ability to mimic human behavior. This approach accurately captures both global and local features of data distributions, leading to comprehensive evaluation conclusions. Results indicate the superiority of our method across all tests.
    \item \textbf{Answers to RQ3}: We proposed an online CAPTCHA bypass framework and evaluation methods. After the real-world experiments, we found that our method significantly outperforms others in pass rates and nearly achieves human-level performance. This demonstrates the high practical value of our approach.
\end{enumerate}

%% file: secs/concul.tex
\section{Conclusion}\label{sec:concul}

Modern CAPTCHAs have the ability to distinguish between human and bot through mouse trajectory analysis, which does not interrupt user's experience. In the pursuit of advancing anti-bot research, we have developed a novel contribution to the field of CAPTCHA generation and evaluation. 
Lacking the capability to exhibit ``controllable randomness," current mouse trajectory generation methods are insufficient in replicating the nuanced variations in human behavior across different testing environments.
We propose a mouse trajectory imitation framework based on the entropy-controlled diffusion model: DMTG. This approach has three main tasks: 1) to generate high simulation mouse trajectories like humans; 2) to evaluate the gap between the generated trajectories and humans using white-box testing; and 3) to deploy and participate in online commercial CAPTCHA black-box evaluations. The experimental results demonstrate that our method minimizes the likelihood of detection of robots and closely mimics human operations in terms of data distribution and real-world evaluations. Besides, we have also identified the reasons behind the effectiveness of our method. These reasons include our ability to copy the speed of mouse movements, including slow initiation and differences in acceleration for different directions. Furthermore, our method has the ability to control curve complexity enabling the rapid elimination of significantly discrepant curves during training. The results from our experiments showcase the versatility of our approach and its potential to provide new directions for enhancing anti-bot measures.

%% file: secs/apedx.tex
\section*{Appendix}

\begin{figure}[t]
    \centering
    \subfigure[DMTG]{
        \includegraphics[width=.45\linewidth]{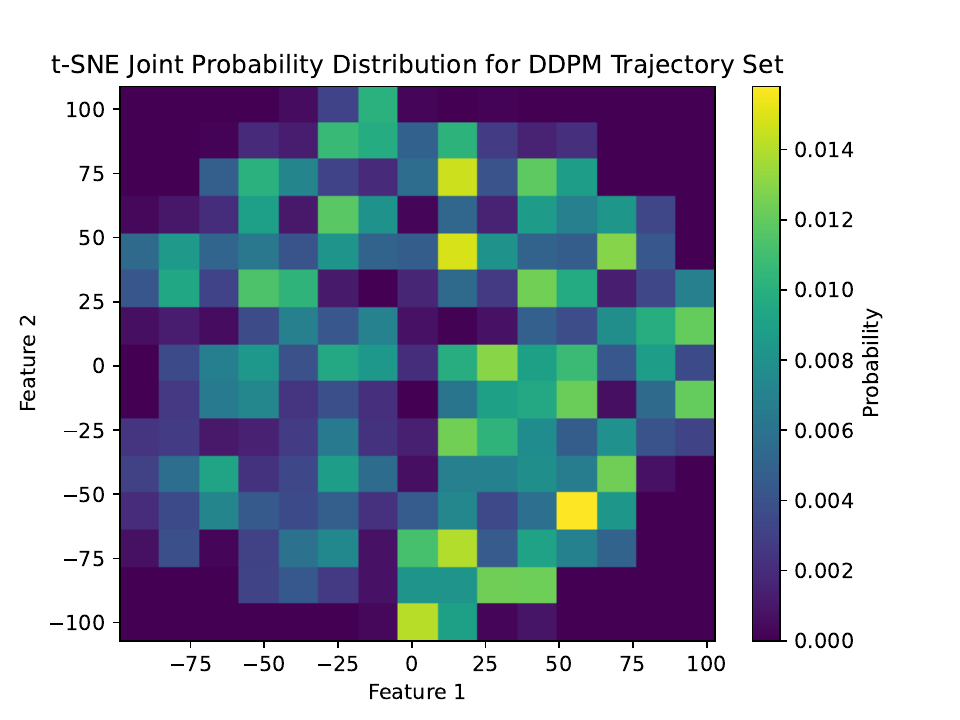}
        \includegraphics[width=.45\linewidth]{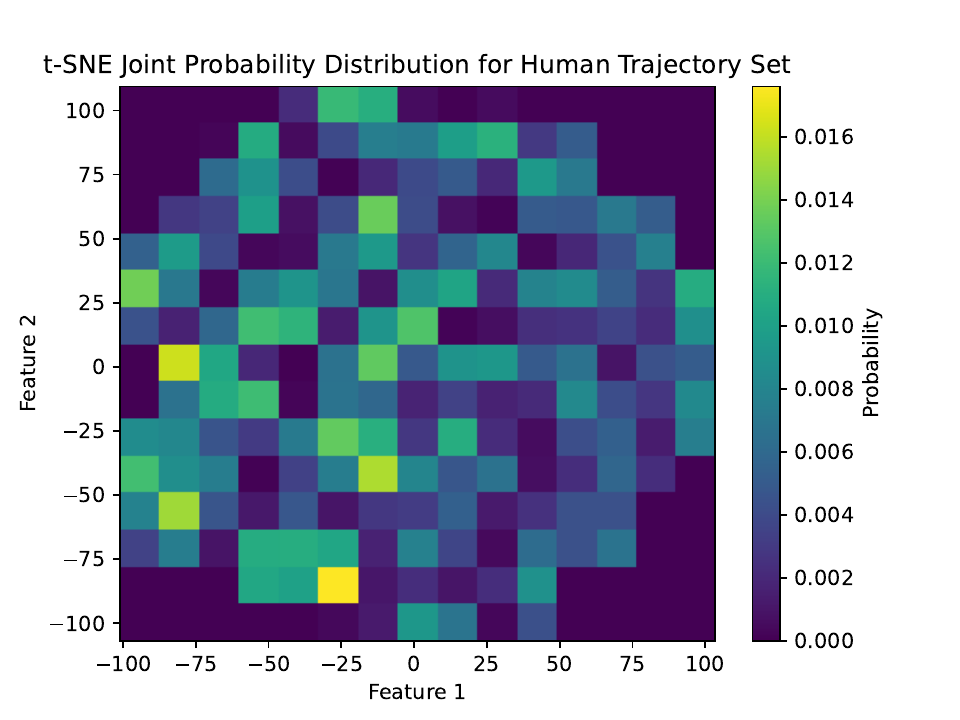}
        \label{fig:DMTG-possibes}
    }
    \subfigure[Ghost]{
        \includegraphics[width=.45\linewidth]{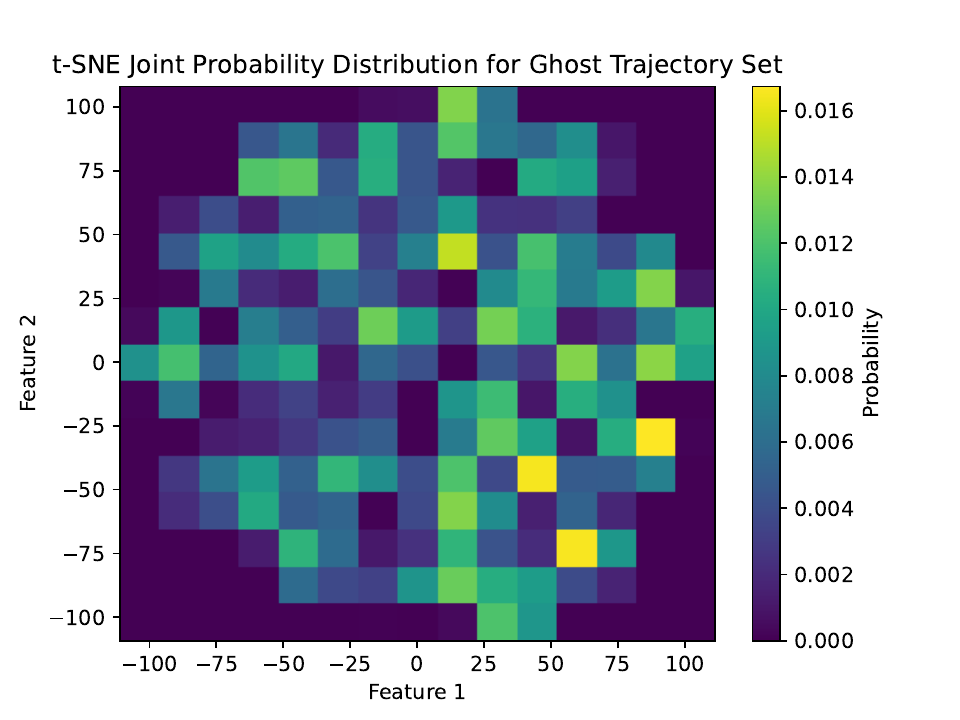}
        \includegraphics[width=.45\linewidth]{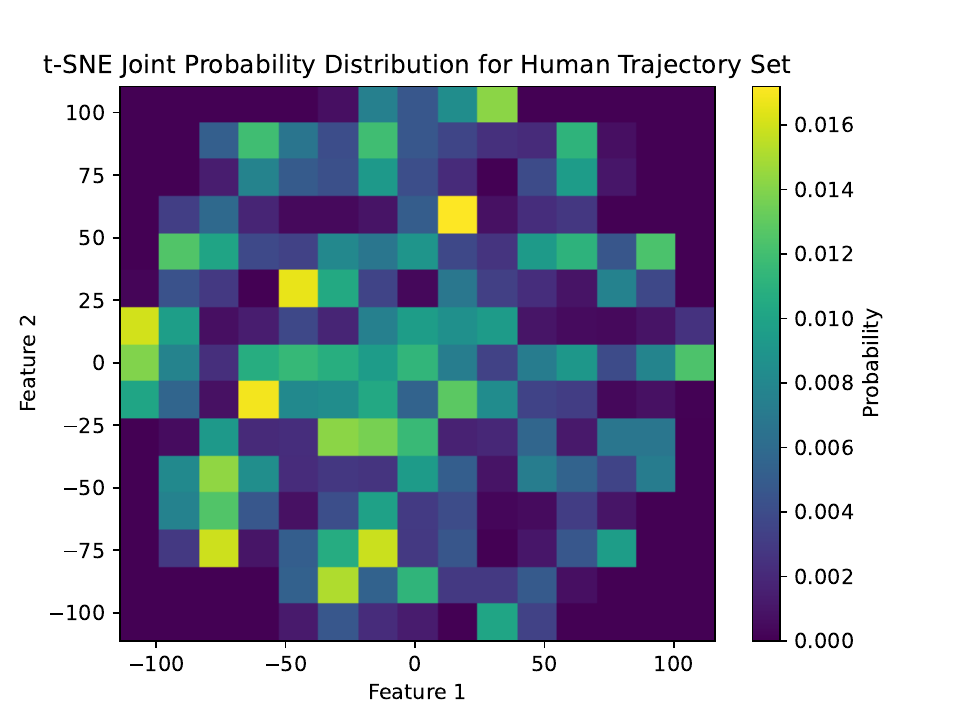}
        \label{fig:ghost-possibes}
    }
    \subfigure[SapiAgent]{
        \includegraphics[width=.45\linewidth]{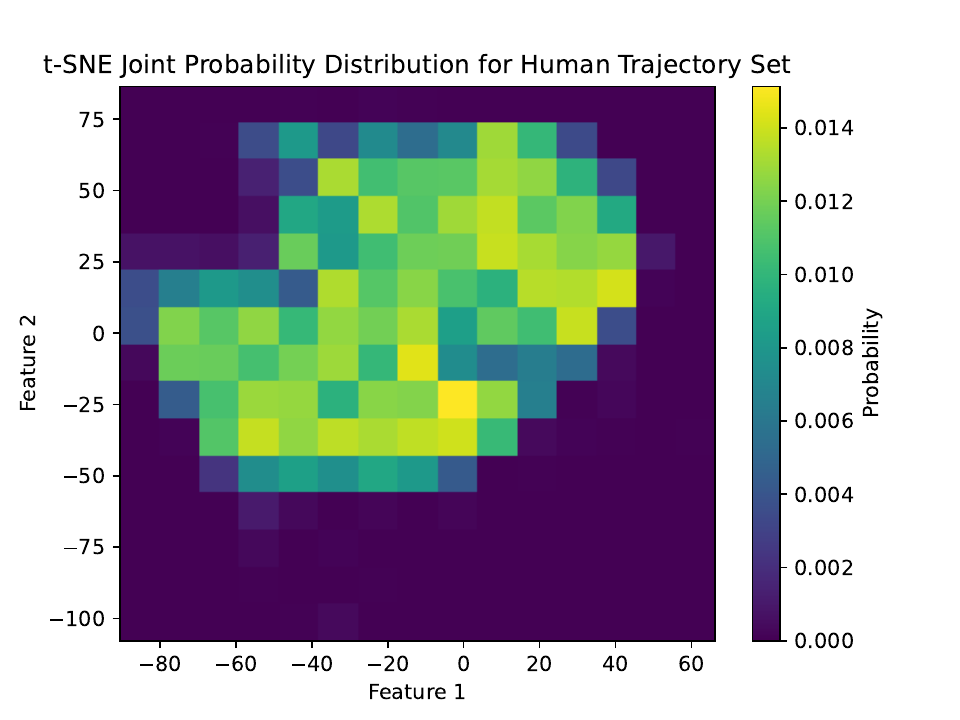}
        \includegraphics[width=.45\linewidth]{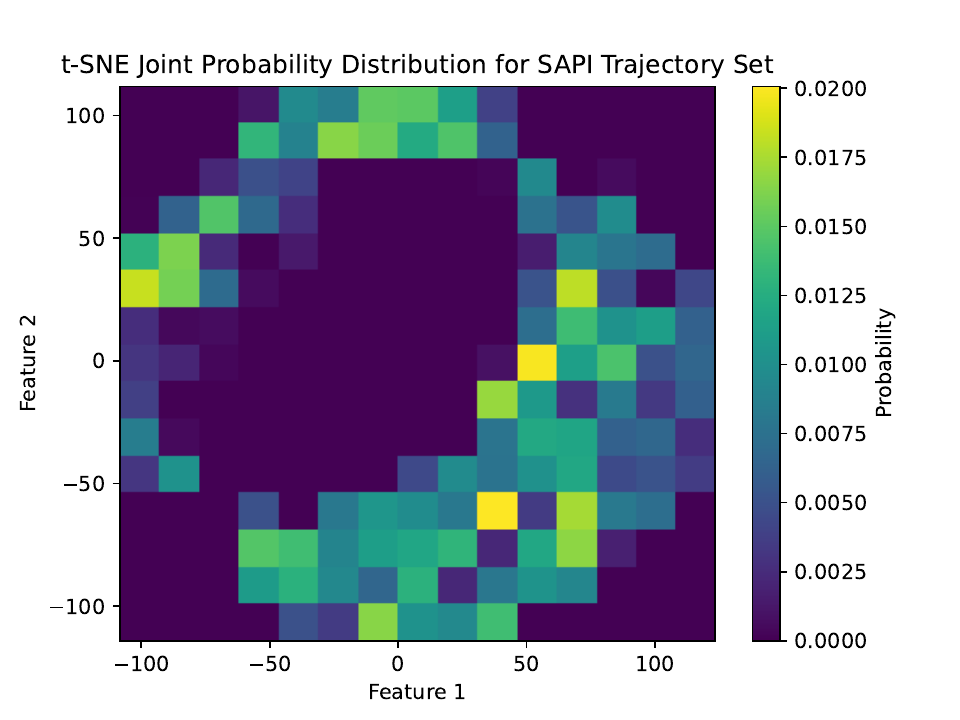}
        \label{fig:sapi-possibes}
    }
    \subfigure[GAN]{
        \includegraphics[width=.45\linewidth]{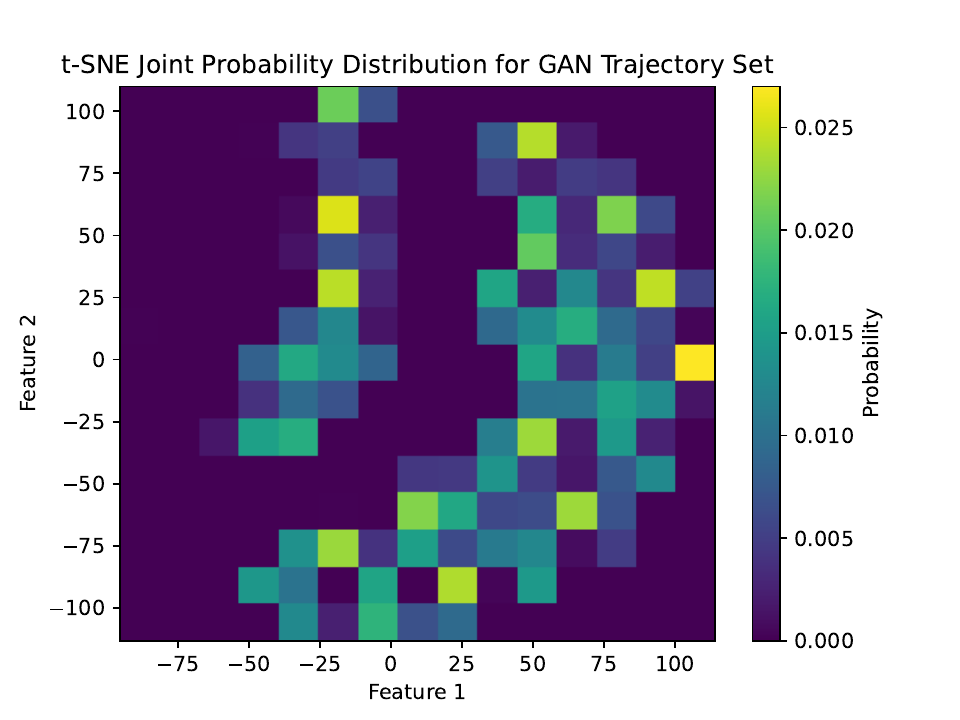}
        \includegraphics[width=.45\linewidth]{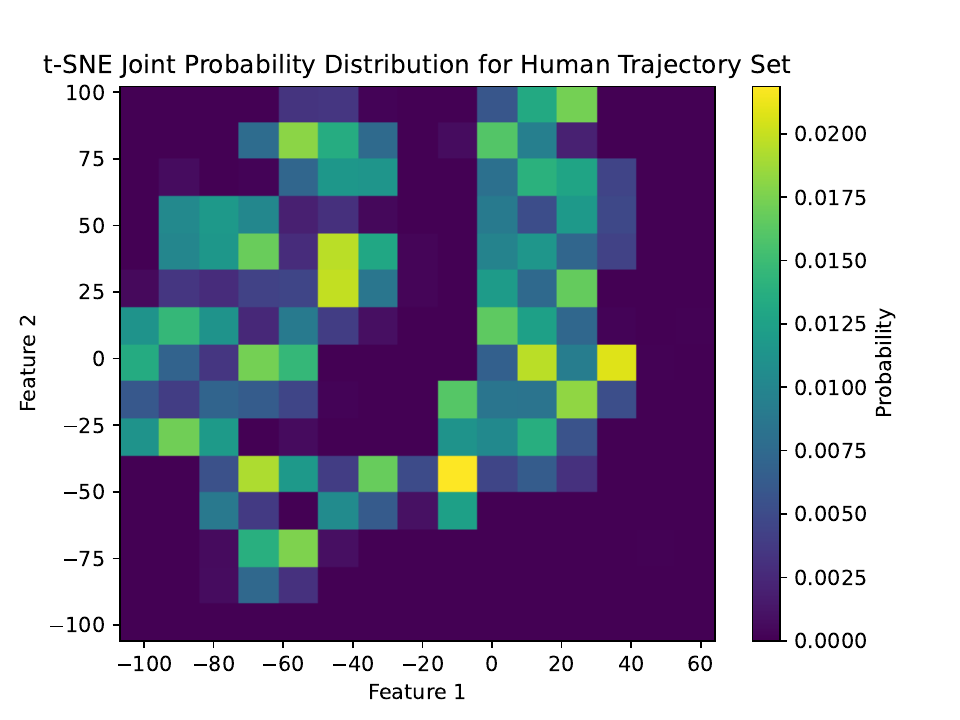}
        \label{fig:gan-possibes}
    }
    \caption{Similarities between Sampling Probability of Human Mouse Trajectories and Models'}
    \label{fig:possibes}
\end{figure}

% \cref{fig:possibes}展示了四种模型在数据分布上和人类的差异，是对\cref{fig:tsne-curves}的补充说明。图左边的是经过t-SNE压缩的模型的数据分布，右边是对应的人类分布。可以发现SapiAgent和GAN几乎与人类的分布没有重合。DMTG分布的空洞少于Ghost，说明在这此实验中，DMTG的采样更均匀。
\cref{fig:possibes} presents the disparities between the four models in data distribution compared to human behavior, serving as supplementary material to \cref{fig:tsne-curves}. On the left side are the t-SNE-compressed data distributions of the models, while on the right side are the corresponding human distributions. It can be observed that SapiAgent and GAN models hardly overlap with the human distribution. DMTG's distribution exhibits fewer voids compared to Ghost, indicating more uniform sampling in this mapping.

% \begin{figure*}[ht]
%     \centering
%     \includegraphics[width=\linewidth]{imgs/DMTG-Proxy.png}
%     \caption{Examples in Fingerprint Resource Pools}
%     \label{fig:DMTG-Proxy}
% \end{figure*}

% % \cref{fig:DMTG-Proxy}展示了我们的网络代理框架中的部分资源的样例。我们通过调取改资源池并配合DMTG完成自动化网络访问。
% \cref{fig:DMTG-Proxy} presents examples of some resources in our browser proxy bot. We utilize this resource pool in conjunction with Entropy-Controlled DDPM to accomplish automated network access.

%% file: main.bbl
\begin{thebibliography}{10}
\providecommand{\url}[1]{#1}
\csname url@samestyle\endcsname
\providecommand{\newblock}{\relax}
\providecommand{\bibinfo}[2]{#2}
\providecommand{\BIBentrySTDinterwordspacing}{\spaceskip=0pt\relax}
\providecommand{\BIBentryALTinterwordstretchfactor}{4}
\providecommand{\BIBentryALTinterwordspacing}{\spaceskip=\fontdimen2\font plus
\BIBentryALTinterwordstretchfactor\fontdimen3\font minus
  \fontdimen4\font\relax}
\providecommand{\BIBforeignlanguage}[2]{{%
\expandafter\ifx\csname l@#1\endcsname\relax
\typeout{** WARNING: IEEEtran.bst: No hyphenation pattern has been}%
\typeout{** loaded for the language `#1'. Using the pattern for}%
\typeout{** the default language instead.}%
\else
\language=\csname l@#1\endcsname
\fi
#2}}
\providecommand{\BIBdecl}{\relax}
\BIBdecl

\bibitem{CAPTCHA}
\BIBentryALTinterwordspacing
X.~Xu, L.~Liu, and B.~Li, ``A survey of captcha technologies to distinguish
  between human and computer,'' \emph{Neurocomputing}, vol. 408, pp. 292--307,
  2020. [Online]. Available:
  \url{https://www.sciencedirect.com/science/article/pii/S0925231220304896}
\BIBentrySTDinterwordspacing

\bibitem{img-captcha-1}
\BIBentryALTinterwordspacing
Y.~Gao, H.~Gao, S.~luo, Y.~Zi, S.~Zhang, W.~Mao, P.~Wang, Y.~Shen, and J.~Yan,
  ``Research on the security of visual reasoning {CAPTCHA},'' in \emph{30th
  USENIX Security Symposium (USENIX Security 21)}.\hskip 1em plus 0.5em minus
  0.4em\relax USENIX Association, Aug. 2021, pp. 3291--3308. [Online].
  Available:
  \url{https://www.usenix.org/conference/usenixsecurity21/presentation/gao}
\BIBentrySTDinterwordspacing

\bibitem{img-captcha-2}
P.~Wang, H.~Gao, C.~Xiao, X.~Guo, Y.~Gao, and Y.~Zi, ``Extended research on the
  security of visual reasoning captcha,'' \emph{IEEE Transactions on Dependable
  and Secure Computing}, vol.~20, no.~6, pp. 4976--4992, 2023.

\bibitem{txt-captcha-1}
\BIBentryALTinterwordspacing
C.~Li, X.~Chen, H.~Wang, P.~Wang, Y.~Zhang, and W.~Wang, ``End-to-end attack on
  text-based captchas based on cycle-consistent generative adversarial
  network,'' \emph{Neurocomputing}, vol. 433, pp. 223--236, 2021. [Online].
  Available:
  \url{https://www.sciencedirect.com/science/article/pii/S0925231220318518}
\BIBentrySTDinterwordspacing

\bibitem{txt-captcha-2}
\BIBentryALTinterwordspacing
P.~Wang, H.~Gao, X.~Guo, C.~Xiao, F.~Qi, and Z.~Yan, ``An experimental
  investigation of text-based captcha attacks and their robustness,'' \emph{ACM
  Comput. Surv.}, vol.~55, no.~9, jan 2023. [Online]. Available:
  \url{https://doi.org/10.1145/3559754}
\BIBentrySTDinterwordspacing

\bibitem{txt-captcha-3}
R.~Zhao, X.~Deng, Y.~Wang, Z.~Yan, Z.~Han, L.~Chen, Z.~Xue, and Y.~Wang,
  ``Geesolver: A generic, efficient, and effortless solver with self-supervised
  learning for breaking text captchas,'' in \emph{2023 IEEE Symposium on
  Security and Privacy (SP)}, 2023, pp. 1649--1666.

\bibitem{YOLO}
T.~Diwan, G.~Anirudh, and J.~V. Tembhurne, ``Object detection using yolo:
  Challenges, architectural successors, datasets and applications,''
  \emph{multimedia Tools and Applications}, vol.~82, no.~6, pp. 9243--9275,
  2023.

\bibitem{llm-captcha-1}
G.~Deng, H.~Ou, Y.~Liu, J.~Zhang, T.~Zhang, and Y.~Liu, ``Oedipus:
  Llm-enchanced reasoning captcha solver,'' \emph{arXiv preprint
  arXiv:2405.07496}, 2024.

\bibitem{mt-becaptcha}
\BIBentryALTinterwordspacing
A.~Acien, A.~Morales, J.~Fierrez, and R.~Vera-Rodriguez, ``Becaptcha-mouse:
  Synthetic mouse trajectories and improved bot detection,'' \emph{Pattern
  Recognition}, vol. 127, p. 108643, 2022. [Online]. Available:
  \url{https://www.sciencedirect.com/science/article/pii/S0031320322001248}
\BIBentrySTDinterwordspacing

\bibitem{mt-captcha-rl-1}
I.~Akrout, A.~Feriani, and M.~Akrout, ``Hacking google recaptcha v3 using
  reinforcement learning,'' \emph{arXiv preprint arXiv:1903.01003}, 2019.

\bibitem{mt-captcha-rl-2}
I.~Tsingenopoulos, D.~Preuveneers, L.~Desmet, and W.~Joosen, ``Captcha me if
  you can: Imitation games with reinforcement learning,'' in \emph{2022 IEEE
  7th European Symposium on Security and Privacy (EuroS\&P)}, 2022, pp.
  719--735.

\bibitem{cc-net}
\BIBentryALTinterwordspacing
P.~C. Humphreys, D.~Raposo, T.~Pohlen, G.~Thornton, R.~Chhaparia, A.~Muldal,
  J.~Abramson, P.~Georgiev, A.~Santoro, and T.~Lillicrap, ``A data-driven
  approach for learning to control computers,'' in \emph{Proceedings of the
  39th International Conference on Machine Learning}, ser. Proceedings of
  Machine Learning Research, K.~Chaudhuri, S.~Jegelka, L.~Song, C.~Szepesvari,
  G.~Niu, and S.~Sabato, Eds., vol. 162.\hskip 1em plus 0.5em minus 0.4em\relax
  PMLR, 17--23 Jul 2022, pp. 9466--9482. [Online]. Available:
  \url{https://proceedings.mlr.press/v162/humphreys22a.html}
\BIBentrySTDinterwordspacing

\bibitem{mt-SapiAgent}
M.~Antal, K.~Buza, and N.~Fejer, ``Sapiagent: A bot based on deep learning to
  generate human-like mouse trajectories,'' \emph{IEEE Access}, vol.~9, pp.
  124\,396--124\,408, 2021.

\bibitem{T-Detector}
S.~Zhao, J.~Fang, S.~Zhao, R.~Wu, J.~Tao, S.~Li, and G.~Pan, ``T-detector: A
  trajectory based pre-trained model for game bot detection in mmorpgs,'' in
  \emph{2022 IEEE 38th International Conference on Data Engineering (ICDE)},
  2022, pp. 992--1003.

\bibitem{DDIM}
\BIBentryALTinterwordspacing
J.~Song, C.~Meng, and S.~Ermon, ``Denoising diffusion implicit models,'' in
  \emph{International Conference on Learning Representations}, 2021. [Online].
  Available: \url{https://openreview.net/forum?id=St1giarCHLP}
\BIBentrySTDinterwordspacing

\bibitem{tencent-vtt}
H.~Wang, F.~Zheng, Z.~Chen, Y.~Lu, J.~Gao, and R.~Wei, ``A captcha design based
  on visual reasoning,'' in \emph{2018 IEEE International Conference on
  Acoustics, Speech and Signal Processing (ICASSP)}, 2018, pp. 1967--1971.

\bibitem{google-recaptcha}
\BIBentryALTinterwordspacing
Google, ``What is recaptcha?'' 2 2024. [Online]. Available:
  \url{https://developers.google.com/recaptcha}
\BIBentrySTDinterwordspacing

\bibitem{audio-captcha-1}
S.~Yilmaz, S.~Zavrak, and H.~Bodur, ``Distinguishing humans from automated
  programs by a novel audio-based captcha,'' \emph{International Journal of
  Computer Applications}, vol. 132, pp. 17--21, 12 2015.

\bibitem{easy-not-secure}
R.~Jin, L.~Huang, J.~Duan, W.~Zhao, Y.~Liao, and P.~Zhou, ``How secure is your
  website? a comprehensive investigation on captcha providers and solving
  services,'' \emph{arXiv preprint arXiv:2306.07543}, 2023.

\bibitem{recaptcha-yolov3}
\BIBentryALTinterwordspacing
M.~I. Hossen, Y.~Tu, M.~F. Rabby, M.~N. Islam, H.~Cao, and X.~Hei, ``An object
  detection based solver for {Google{\textquoteright}s} image {reCAPTCHA} v2,''
  in \emph{23rd International Symposium on Research in Attacks, Intrusions and
  Defenses (RAID 2020)}.\hskip 1em plus 0.5em minus 0.4em\relax San Sebastian:
  USENIX Association, Oct. 2020, pp. 269--284. [Online]. Available:
  \url{https://www.usenix.org/conference/raid2020/presentation/hossen}
\BIBentrySTDinterwordspacing

\bibitem{yolov3}
J.~Redmon and A.~Farhadi, ``Yolov3: An incremental improvement,'' \emph{arXiv
  preprint arXiv:1804.02767}, 2018.

\bibitem{txt-rcnn-att-chinese}
P.~Wang, H.~Gao, Q.~Rao, S.~Luo, Z.~Yuan, and Z.~Shi, ``A security analysis of
  captchas with large character sets,'' \emph{IEEE Transactions on Dependable
  and Secure Computing}, vol.~18, no.~6, pp. 2953--2968, 2021.

\bibitem{RNN}
M.~Schuster and K.~Paliwal, ``Bidirectional recurrent neural networks,''
  \emph{IEEE Transactions on Signal Processing}, vol.~45, no.~11, pp.
  2673--2681, 1997.

\bibitem{CNN}
Z.~Li, F.~Liu, W.~Yang, S.~Peng, and J.~Zhou, ``A survey of convolutional
  neural networks: Analysis, applications, and prospects,'' \emph{IEEE
  Transactions on Neural Networks and Learning Systems}, vol.~33, no.~12, pp.
  6999--7019, 2022.

\bibitem{ATT}
A.~Vaswani, N.~Shazeer, N.~Parmar, J.~Uszkoreit, L.~Jones, A.~N. Gomez, L.~u.
  Kaiser, and I.~Polosukhin, ``Attention is all you need,'' in \emph{Advances
  in Neural Information Processing Systems}, I.~Guyon, U.~V. Luxburg,
  S.~Bengio, H.~Wallach, R.~Fergus, S.~Vishwanathan, and R.~Garnett, Eds.,
  vol.~30.\hskip 1em plus 0.5em minus 0.4em\relax Curran Associates, Inc.,
  2017.

\bibitem{Mouse2Vec}
\BIBentryALTinterwordspacing
G.~Zhang, Z.~Hu, M.~B\^{a}ce, and A.~Bulling, ``Mouse2vec: Learning reusable
  semantic representations of mouse behaviour,'' in \emph{Proceedings of the
  CHI Conference on Human Factors in Computing Systems}, ser. CHI '24.\hskip
  1em plus 0.5em minus 0.4em\relax New York, NY, USA: Association for Computing
  Machinery, 2024. [Online]. Available:
  \url{https://doi.org/10.1145/3613904.3642141}
\BIBentrySTDinterwordspacing

\bibitem{ReMouse}
\BIBentryALTinterwordspacing
S.~Sadeghpour and N.~Vlajic, ``Remouse dataset: On the efficacy of measuring
  the similarity of human-generated trajectories for the detection of
  session-replay bots,'' \emph{Journal of Cybersecurity and Privacy}, vol.~3,
  no.~1, pp. 95--117, 2023. [Online]. Available:
  \url{https://www.mdpi.com/2624-800X/3/1/7}
\BIBentrySTDinterwordspacing

\bibitem{phone-traj}
A.~Acien, A.~Morales, R.~Vera-Rodriguez, and J.~Fierrez, ``Smartphone sensors
  for modeling human-computer interaction: General outlook and research
  datasets for user authentication,'' in \emph{2020 IEEE 44th Annual Computers,
  Software, and Applications Conference (COMPSAC)}, 2020, pp. 1273--1278.

\bibitem{jin-mouse-traj}
R.~Jin, Y.~Liao, and P.~Zhou, ``User authentication and identity inconsistency
  detection via mouse-trajectory similarity measurement,'' \emph{arXiv preprint
  arXiv:2312.10273}, 2023.

\bibitem{GAN2}
I.~Goodfellow, J.~Pouget-Abadie, M.~Mirza, B.~Xu, D.~Warde-Farley, S.~Ozair,
  A.~Courville, and Y.~Bengio, ``Generative adversarial nets,'' in
  \emph{Advances in Neural Information Processing Systems}, Z.~Ghahramani,
  M.~Welling, C.~Cortes, N.~Lawrence, and K.~Weinberger, Eds., vol.~27.\hskip
  1em plus 0.5em minus 0.4em\relax Curran Associates, Inc., 2014.

\bibitem{mt-captcha-3}
S.~E. Folch, A.~C. Ibáñez, N.~O. Rabella, and J.~E. Escrig, ``Web bot
  detection using mouse movement,'' in \emph{2023 JNIC Cybersecurity Conference
  (JNIC)}, 2023, pp. 1--6.

\bibitem{MiniWob}
E.~Z. Liu, K.~Guu, P.~Pasupat, T.~Shi, and P.~Liang, ``Reinforcement learning
  on web interfaces using workflow-guided exploration,'' \emph{arXiv preprint
  arXiv:1802.08802}, 2018.

\bibitem{RL}
L.~P. Kaelbling, M.~L. Littman, and A.~W. Moore, ``Reinforcement learning: A
  survey,'' \emph{Journal of artificial intelligence research}, vol.~4, pp.
  237--285, 1996.

\bibitem{GAI-product-effect}
\BIBentryALTinterwordspacing
S.~Noy and W.~Zhang, ``Experimental evidence on the productivity effects of
  generative artificial intelligence,'' \emph{Science}, vol. 381, no. 6654, pp.
  187--192, 2023. [Online]. Available:
  \url{https://www.science.org/doi/abs/10.1126/science.adh2586}
\BIBentrySTDinterwordspacing

\bibitem{GAI-defination}
D.~Baidoo-anu and L.~Owusu~Ansah, ``Education in the era of generative
  artificial intelligence (ai): Understanding the potential benefits of chatgpt
  in promoting teaching and learning,'' \emph{Journal of AI}, vol.~7, no.~1, p.
  52–62, 2023.

\bibitem{AutoEncoder}
\BIBentryALTinterwordspacing
C.-Y. Liou, W.-C. Cheng, J.-W. Liou, and D.-R. Liou, ``Autoencoder for words,''
  \emph{Neurocomputing}, vol. 139, pp. 84--96, 2014. [Online]. Available:
  \url{https://www.sciencedirect.com/science/article/pii/S0925231214003658}
\BIBentrySTDinterwordspacing

\bibitem{VAE}
D.~P. Kingma and M.~Welling, ``Auto-encoding variational bayes,'' \emph{arXiv
  preprint arXiv:1312.6114}, 2013.

\bibitem{DAE}
\BIBentryALTinterwordspacing
P.~Vincent, H.~Larochelle, Y.~Bengio, and P.-A. Manzagol, ``Extracting and
  composing robust features with denoising autoencoders,'' in \emph{Proceedings
  of the 25th International Conference on Machine Learning}, ser. ICML
  '08.\hskip 1em plus 0.5em minus 0.4em\relax New York, NY, USA: Association
  for Computing Machinery, 2008, p. 1096–1103. [Online]. Available:
  \url{https://doi.org/10.1145/1390156.1390294}
\BIBentrySTDinterwordspacing

\bibitem{OpenVoice}
Z.~Qin, W.~Zhao, X.~Yu, and X.~Sun, ``Openvoice: Versatile instant voice
  cloning,'' \emph{arXiv preprint arXiv:2312.01479}, 2023.

\bibitem{GAN}
\BIBentryALTinterwordspacing
I.~Goodfellow, J.~Pouget-Abadie, M.~Mirza, B.~Xu, D.~Warde-Farley, S.~Ozair,
  A.~Courville, and Y.~Bengio, ``Generative adversarial networks,''
  \emph{Commun. ACM}, vol.~63, no.~11, p. 139–144, oct 2020. [Online].
  Available: \url{https://doi.org/10.1145/3422622}
\BIBentrySTDinterwordspacing

\bibitem{CycleGAN}
J.-Y. Zhu, T.~Park, P.~Isola, and A.~A. Efros, ``Unpaired image-to-image
  translation using cycle-consistent adversarial networks,'' in
  \emph{Proceedings of the IEEE International Conference on Computer Vision
  (ICCV)}, 2017, pp. 2223--2232.

\bibitem{StyleGAN}
T.~Karras, S.~Laine, and T.~Aila, ``A style-based generator architecture for
  generative adversarial networks,'' in \emph{Proceedings of the IEEE/CVF
  conference on computer vision and pattern recognition}, 2019, pp. 4401--4410.

\bibitem{ViT}
K.~Han, Y.~Wang, H.~Chen, X.~Chen, J.~Guo, Z.~Liu, Y.~Tang, A.~Xiao, C.~Xu,
  Y.~Xu, Z.~Yang, Y.~Zhang, and D.~Tao, ``A survey on vision transformer,''
  \emph{IEEE Transactions on Pattern Analysis and Machine Intelligence},
  vol.~45, no.~1, pp. 87--110, 2023.

\bibitem{GPT}
T.~Brown, B.~Mann, N.~Ryder, M.~Subbiah, J.~D. Kaplan, P.~Dhariwal,
  A.~Neelakantan, P.~Shyam, G.~Sastry, A.~Askell \emph{et~al.}, ``Language
  models are few-shot learners,'' \emph{Advances in neural information
  processing systems}, vol.~33, pp. 1877--1901, 2020.

\bibitem{Dall-E}
\BIBentryALTinterwordspacing
A.~Ramesh, M.~Pavlov, G.~Goh, S.~Gray, C.~Voss, A.~Radford, M.~Chen, and
  I.~Sutskever, ``Zero-shot text-to-image generation,'' in \emph{Proceedings of
  the 38th International Conference on Machine Learning}, ser. Proceedings of
  Machine Learning Research, M.~Meila and T.~Zhang, Eds., vol. 139.\hskip 1em
  plus 0.5em minus 0.4em\relax PMLR, 18--24 Jul 2021, pp. 8821--8831. [Online].
  Available: \url{https://proceedings.mlr.press/v139/ramesh21a.html}
\BIBentrySTDinterwordspacing

\bibitem{DiffusionNet}
\BIBentryALTinterwordspacing
J.~Sohl-Dickstein, E.~Weiss, N.~Maheswaranathan, and S.~Ganguli, ``Deep
  unsupervised learning using nonequilibrium thermodynamics,'' in
  \emph{Proceedings of the 32nd International Conference on Machine Learning},
  ser. Proceedings of Machine Learning Research, F.~Bach and D.~Blei, Eds.,
  vol.~37.\hskip 1em plus 0.5em minus 0.4em\relax Lille, France: PMLR, 07--09
  Jul 2015, pp. 2256--2265. [Online]. Available:
  \url{https://proceedings.mlr.press/v37/sohl-dickstein15.html}
\BIBentrySTDinterwordspacing

\bibitem{DDPM}
J.~Ho, A.~Jain, and P.~Abbeel, ``Denoising diffusion probabilistic models,'' in
  \emph{Advances in Neural Information Processing Systems}, H.~Larochelle,
  M.~Ranzato, R.~Hadsell, M.~Balcan, and H.~Lin, Eds., vol.~33.\hskip 1em plus
  0.5em minus 0.4em\relax Curran Associates, Inc., 2020, pp. 6840--6851.

\bibitem{AlphaFold3}
\BIBentryALTinterwordspacing
J.~Abramson, J.~Adler, J.~Dunger, R.~Evans, T.~Green, A.~Pritzel,
  O.~Ronneberger, L.~Willmore, A.~J. Ballard, J.~Bambrick \emph{et~al.},
  ``Accurate structure prediction of biomolecular interactions with alphafold
  3,'' \emph{Nature}, pp. 1--3, 5 2024. [Online]. Available:
  \url{https://doi.org/10.1038/s41586-024-07487-w}
\BIBentrySTDinterwordspacing

\bibitem{llama}
H.~Touvron, T.~Lavril, G.~Izacard, X.~Martinet, M.-A. Lachaux, T.~Lacroix,
  B.~Rozi{\`e}re, N.~Goyal, E.~Hambro, F.~Azhar \emph{et~al.}, ``Llama: Open
  and efficient foundation language models,'' \emph{arXiv preprint
  arXiv:2302.13971}, 2023.

\bibitem{vit-gan-survey}
S.~R. Dubey and S.~K. Singh, ``Transformer-based generative adversarial
  networks in computer vision: A comprehensive survey,'' \emph{IEEE
  Transactions on Artificial Intelligence}, pp. 1--16, 2024.

\bibitem{vit-survey}
L.~Papa, P.~Russo, I.~Amerini, and L.~Zhou, ``A survey on efficient vision
  transformers: Algorithms, techniques, and performance benchmarking,''
  \emph{IEEE Transactions on Pattern Analysis and Machine Intelligence}, pp.
  1--20, 2024.

\bibitem{MedSegDiff}
\BIBentryALTinterwordspacing
J.~Wu, W.~Ji, H.~Fu, M.~Xu, Y.~Jin, and Y.~Xu, ``Medsegdiff-v2: Diffusion-based
  medical image segmentation with transformer,'' in \emph{Proceedings of the
  AAAI Conference on Artificial Intelligence}, vol.~38, no.~6, 3 2024, pp.
  6030--6038. [Online]. Available:
  \url{https://ojs.aaai.org/index.php/AAAI/article/view/28418}
\BIBentrySTDinterwordspacing

\bibitem{u-net}
O.~Ronneberger, P.~Fischer, and T.~Brox, ``U-net: Convolutional networks for
  biomedical image segmentation,'' \emph{arXiv preprint arXiv:1505.04597},
  2015.

\bibitem{UniDiffuser}
\BIBentryALTinterwordspacing
F.~Bao, S.~Nie, K.~Xue, C.~Li, S.~Pu, Y.~Wang, G.~Yue, Y.~Cao, H.~Su, and
  J.~Zhu, ``One transformer fits all distributions in multi-modal diffusion at
  scale,'' in \emph{Proceedings of the 40th International Conference on Machine
  Learning}, ser. Proceedings of Machine Learning Research, A.~Krause,
  E.~Brunskill, K.~Cho, B.~Engelhardt, S.~Sabato, and J.~Scarlett, Eds., vol.
  202.\hskip 1em plus 0.5em minus 0.4em\relax PMLR, 23--29 Jul 2023, pp.
  1692--1717. [Online]. Available:
  \url{https://proceedings.mlr.press/v202/bao23a.html}
\BIBentrySTDinterwordspacing

\bibitem{difu-survey}
H.~Cao, C.~Tan, Z.~Gao, Y.~Xu, G.~Chen, P.-A. Heng, and S.~Z. Li, ``A survey on
  generative diffusion models,'' \emph{IEEE Transactions on Knowledge and Data
  Engineering}, pp. 1--20, 2024.

\bibitem{v-difu-survey}
F.-A. Croitoru, V.~Hondru, R.~T. Ionescu, and M.~Shah, ``Diffusion models in
  vision: A survey,'' \emph{IEEE Transactions on Pattern Analysis and Machine
  Intelligence}, vol.~45, no.~9, pp. 10\,850--10\,869, 2023.

\bibitem{difu-survey-2}
\BIBentryALTinterwordspacing
L.~Yang, Z.~Zhang, Y.~Song, S.~Hong, R.~Xu, Y.~Zhao, W.~Zhang, B.~Cui, and
  M.-H. Yang, ``Diffusion models: A comprehensive survey of methods and
  applications,'' \emph{ACM Comput. Surv.}, vol.~56, no.~4, nov 2023. [Online].
  Available: \url{https://doi.org/10.1145/3626235}
\BIBentrySTDinterwordspacing

\bibitem{SapiMouse}
M.~Antal, N.~Fejér, and K.~Buza, ``Sapimouse: Mouse dynamics-based user
  authentication using deep feature learning,'' in \emph{2021 IEEE 15th
  International Symposium on Applied Computational Intelligence and Informatics
  (SACI)}, 2021, pp. 61--66.

\bibitem{openimages}
A.~Kuznetsova, H.~Rom, N.~Alldrin, J.~Uijlings, I.~Krasin, J.~Pont-Tuset,
  S.~Kamali, S.~Popov, M.~Malloci, A.~Kolesnikov \emph{et~al.}, ``The open
  images dataset v4: Unified image classification, object detection, and visual
  relationship detection at scale,'' \emph{International journal of computer
  vision}, vol. 128, no.~7, pp. 1956--1981, 2020.

\bibitem{jsd}
\BIBentryALTinterwordspacing
M.~Menéndez, J.~Pardo, L.~Pardo, and M.~Pardo, ``The jensen-shannon
  divergence,'' \emph{Journal of the Franklin Institute}, vol. 334, no.~2, pp.
  307--318, 1997. [Online]. Available:
  \url{https://www.sciencedirect.com/science/article/pii/S0016003296000634}
\BIBentrySTDinterwordspacing

\bibitem{Wasserstein}
\BIBentryALTinterwordspacing
V.~M. Panaretos and Y.~Zemel, ``Statistical aspects of wasserstein distances,''
  \emph{Annual Review of Statistics and Its Application}, vol.~6, pp. 405--431,
  2019. [Online]. Available:
  \url{https://www.annualreviews.org/content/journals/10.1146/annurev-statistics-030718-104938}
\BIBentrySTDinterwordspacing

\bibitem{emd}
Y.~Rubner, C.~Tomasi, and L.~J. Guibas, ``The earth mover's distance as a
  metric for image retrieval,'' \emph{International journal of computer
  vision}, vol.~40, pp. 99--121, 2000.

\bibitem{tsne}
\BIBentryALTinterwordspacing
L.~van~der Maaten and G.~Hinton, ``Visualizing data using t-sne,''
  \emph{Journal of Machine Learning Research}, vol.~9, no.~86, pp. 2579--2605,
  2008. [Online]. Available:
  \url{http://jmlr.org/papers/v9/vandermaaten08a.html}
\BIBentrySTDinterwordspacing

\bibitem{Akamai}
\BIBentryALTinterwordspacing
Akamai, ``Power and protect life online,'' 2024. [Online]. Available:
  \url{https://www.akamai.com/why-akamai}
\BIBentrySTDinterwordspacing

\bibitem{Geetest}
\BIBentryALTinterwordspacing
GeeTest, ``Geetest captcha v3,'' 2024. [Online]. Available:
  \url{https://www.geetest.com/en/Captcha}
\BIBentrySTDinterwordspacing

\bibitem{bezier}
\BIBentryALTinterwordspacing
H.~Prautzsch, W.~Boehm, and M.~Paluszny, \emph{B{\'e}zier and B-spline
  techniques}.\hskip 1em plus 0.5em minus 0.4em\relax Springer, 2002, vol.~6.
  [Online]. Available: \url{https://doi.org/10.1007/978-3-662-04919-8}
\BIBentrySTDinterwordspacing

\bibitem{ghost}
\BIBentryALTinterwordspacing
Xetera, ``Ghost cursor,'' 2024. [Online]. Available:
  \url{https://github.com/Xetera/ghost-cursor}
\BIBentrySTDinterwordspacing

\bibitem{dt}
Y.-Y. Song and L.~Ying, ``Decision tree methods: applications for
  classification and prediction,'' \emph{Shanghai archives of psychiatry},
  vol.~27, no.~2, p. 130–135, 2015.

\bibitem{rf}
L.~Breiman, ``Random forests,'' \emph{Machine learning}, vol.~45, pp. 5--32,
  2001.

\bibitem{XGBoost}
A.~Ogunleye and Q.-G. Wang, ``Xgboost model for chronic kidney disease
  diagnosis,'' \emph{IEEE/ACM Transactions on Computational Biology and
  Bioinformatics}, vol.~17, no.~6, pp. 2131--2140, 2020.

\bibitem{graboost}
C.~Bent{\'e}jac, A.~Cs{\"o}rg{\H{o}}, and G.~Mart{\'\i}nez-Mu{\~n}oz, ``A
  comparative analysis of gradient boosting algorithms,'' \emph{Artificial
  Intelligence Review}, vol.~54, pp. 1937--1967, 2021.

\bibitem{MLP}
A.~Pinkus, ``Approximation theory of the mlp model in neural networks,''
  \emph{Acta Numerica}, vol.~8, p. 143–195, 1999.

\bibitem{LSTM}
\BIBentryALTinterwordspacing
A.~Graves, \emph{Long Short-Term Memory}.\hskip 1em plus 0.5em minus
  0.4em\relax Berlin, Heidelberg: Springer Berlin Heidelberg, 2012, pp. 37--45.
  [Online]. Available: \url{https://doi.org/10.1007/978-3-642-24797-2_4}
\BIBentrySTDinterwordspacing

\bibitem{BiLSTM}
S.~Zhang, D.~Zheng, X.~Hu, and M.~Yang, ``Bidirectional long short-term memory
  networks for relation classification,'' in \emph{Proceedings of the 29th
  Pacific Asia conference on language, information and computation}, 2015, pp.
  73--78.

\bibitem{TCN}
S.~Bai, J.~Z. Kolter, and V.~Koltun, ``An empirical evaluation of generic
  convolutional and recurrent networks for sequence modeling,'' \emph{arXiv
  preprint arXiv:1803.01271}, 2018.

\end{thebibliography}
